# The Qatar Genome: A Population-Specific Tool for Precision Medicine in the Middle East*

Khalid A. Fakhro[1,2], Michelle R. Staudt[3], Monica Denise Ramstetter[3], Amal Robay[2], Joel A. Malek[2], Ramin Badii[4], Ajayeb Al-Nabet Al-Marri[4], Charbel Abi Khalil[2], Alya Al-Shakaki[2], Omar Chidiac[2], Dora Stadler[5], Mahmoud Zirie[6], Amin Jayyousi[6], Jacqueline Salit[3], Jason G. Mezey[3,7], Ronald G. Crystal[3], and Juan L. Rodriguez-Flores[3]

[1] Sidra Medical and Research Center,
Doha, Qatar,

[2]Department of Genetic Medicine
[5]Weill Cornell Medical College in Qatar
Doha, Qatar

[3]Department of Genetic Medicine
Weill Cornell Medical College
New York, New York,

[4]Laboratory Medicine and Pathology
[6]Department of Medicine
Hamad Medical Corporation
Doha, Qatar

and

[7]Department of Biological Statistics and Computational Biology
Cornell University
Ithaca, NY

* These studies were supported, in part, by the Qatar Foundation and Weill Cornell Medical College in Qatar; and NIH T32 HL09428; Qatar National Research Fund NPRP 5-436-3-116 and NPRP 7-1425-3-370.

Running head:    Qatari genome

Correspondence:    Department of Genetic Medicine
Weill Cornell Medical College
1300 York Avenue, Box 164
New York, New York 10065
Phone:   (646) 962-4363
Fax:     (646) 962-0220
E-mail:  geneticmedicine@med.cornell.edu

**Abstract**

Reaching the full potential of precision medicine depends on the quality of personalized genome interpretation. In order to facilitate precision medicine in regions of the Middle East and North Africa (MENA), a population-specific reference genome for the indigenous Arab population of Qatar (QTRG) was constructed by incorporating allele frequency data from sequencing of 1,161 Qataris, representing 0.4% of the population. A total of 20.9 million SNP and 3.1 million indels were observed in Qatar, including an average of 1.79% novel variants per individual genome. Replacement of the GRCh37 standard reference with QTRG in a best practices genome analysis workflow resulted in an average of 7* deeper coverage depth (an improvement of 23%), and 756,671 fewer variants on average, a reduction of 16% that is attributed to common Qatari alleles being present in the QTRG reference. The benefit for using QTRG varies across ancestries, a factor that should be taken into consideration when selecting an appropriate reference for analysis.

**Introduction**

Precision medicine involves tailoring medical decision making to genomic individuality in the context of an individual's unique environment/lifestyle[1]. Early examples of successful application of precision medicine include cancer, where sequencing of the patient and tumor genomes can identify specific targets for therapeutic decisions[2], and rare disease, where sequencing can lead to the rapid discovery of causative mutations and correct diagnosis in a time-critical clinical setting[3]. For the latter, while the cost of sequencing an individual's genome has declined rapidly in the past 1.5 decades, there are still considerable challenges for genome interpretation, such as important variants being missed due to low coverage or incorrect calls, and the emerging challenge of interpreting a growing number of variants of unknown significance in each human genome. With over 3 million SNPs identified per human genome sequenced, the majority of these variants cannot be immediately linked to a known phenotype, and the putative impact of variants must therefore be inferred computationally using algorithms that harness comparative genomics and available experimental data[4-6]. Thus, precision medicine in the near term stands to benefit greatly from both increased accuracy in variant calling and improved interpretability when the aim is to identify variants of relevance to one or more phenotypic manifestations within an individual, family of related individuals, or population.

Reference bias is a known issue in human genome resequencing for variant detection[7], and modifications to the reference can improve calling accuracy and interpretability[8]. Relevant to the issue of variant calling accuracy, a reference that more closely matches the ancestry of the genome(s) being aligned is expected to reduce mismatches during alignment and lead to more accurate genotypes[8]. Applicable to the issue of interpretability of variants for rare disease is the observation that a variant's prevalence is inversely proportional to the variant's deleterious impact on cellular function (and by extension, evolutionary fitness), with the most severely delete-rious variants being the rarest due to purifying selection[9]. Based on this principle, the American College of Medical Genetics (ACMG) recommends excluding any allele above 5% prevalence from consideration as pathogenic[10]. Given that allele frequency is a population-dependent property (e.g., an allele that is rare in one population may be common in another), relying entirely on the standard reference genome (GRCh37) or allele frequency in ethnically mismatched populations (even if a large sample of individual genomes has been sequenced) may produce incorrect assessments of the pathogenicity of a specific allele in an under-studied population.

One approach to improving both variant calling accuracy and interpretability of an individual's genome is to incorporate variant prevalence information early on in the genome interpretation process by modifying the reference genome, such that variants discovered in the genome are the minor allele in the population[8]. This modification to the reference results in a streamlined analysis workflow, as fewer variants need to be interpreted. In this context, it should be of value to produce a separate major allele reference genome for each distinct ancestral population or regional meta-population, particularly in cases when the genomic variation in these populations has not been well sampled in public databases.

The region of the Middle East and North Africa (MENA) is an example of such an under-studied meta-population where the rising adoption of precision medicine could benefit greatly from a specific major allele reference genome. In particular, this region is characterized by a high prevalence of consanguineous marriage and elevated incidence of certain Mendelian disorders, and thus is a region where numerous disease studies are faced with the problem of too many potential disease variants per sequencing experiment[11]. Such a region would undoubtedly benefit from allele frequency databases of ethnically matched controls. In addition, given the diversity of the region, many variants called in the population relative to the current reference genome may in fact be the major allele in the population, and eliminating these from being called would lead to a more efficient analysis.

To produce a first version of a reference genome tailored to the region, we sampled and sequenced genomes from Qataris, an indigenous population located near the center of the MENA region. The current population of the nation of Qatar includes over 1.7 million expatriates primarily from MENA and South Asia that have arrived in recent decades[12], and ~300,000 Qataris of indigenous ancestry who arrived in prior waves of migration. At the genome level, the ethnic Qataris represent a population with a mixture of Bedouin/Arab (Q1), Persian/South Asian (Q2) and African (Q3) ancestry[13,14]. In particular, the Bedouin/Arab (Q1) subpopulation, with deepest ancestral roots in the Peninsula, continue to practice within-tribe marriage that has led to a high level of homozygosity compared to other populations worldwide[14,15].

In order to demonstrate the value of a reference genome tailored to an indigenous population in the MENA region, we have constructed a Qatari Reference Genome (QTRG) where the reference bases are "flipped" to the Qatari major allele. The value and utility of QTRG with respect to the standard reference (GRCh37) was evaluated in several ways on genomes not used to construct the QTRG. In the field of population genomics, the term "n+1 genome" refers to the "next" genome sequenced after a large-scale sequencing effort of n genomes, such as the 1000 Genomes Project. A major question in population genomics is "what is the benefit of having a database of n sequenced genomes when sequencing a single genome not in the database?", where the single genome is referred to as the "n+1" genome. In this study, the value of the database of over 1000 sequenced Qatari genomes and exomes is demonstrated, including analysis of over n=15 genomes and n=16 exomes from diverse ancestries. Improvements in mapped read depth were observed, and the subsequent improvement in variant sensitivity was measured. A catalog of known pathogenic variants in Qatar was compiled, with variant coordinates in both the standard (GRCh37) and modified (QTRG) reference.

## Methods Summary

Human subjects were recruited and written informed consent was obtained at Hamad Medical Corporation (HMC) and HMC Primary Health Care Centers in Doha, Qatar under protocols approved by the Institutional Review Boards of Hamad Medical Corporation and Weill Cornell Medical College in Qatar. Briefly, a total of 1,376 subjects were recruited for genome (n=108) or exome (n=1,268) sequencing, including a set of 31 sequenced by both methods for validation purposes. The exome sequencing included target enrichment using either the Agilent SureSelect Human All Exon 38 Mb (n=67) (referred to as Exome38Mb) and Agilent SureSelect Human All Exon 51 Mb (n=1201) platforms (referred to as Exome51Mb). Subjects included both 3-generation Qataris (n=1,161) and non-Qatari residents of Qatar (n=215), from the general Middle East and North Africa (MENA) region or South Asia. Genotypes were generated using the GATK Best Practices workflow[16]. One Qatari female was sequenced on all three platforms, as well as a fourth platform (Illumina HiSeq X), and was used for calibration of batch-specific filters for data integration. In order to minimize batch-specific variants, a range of batch-specific filters were evaluated on genomic intervals in the intersection of the three platforms, and the optimal minimum depth and minimum allele count filters were selected, such that the novel SNP rate was consistent across batches within genomic intervals covered by the four platforms (0.73% novel SNPs), resulting in under 5% batch-specific variants in the quadruple-sequenced Qatari. The filters were confirmed to not be overly stringent by assessing the number of coding variants (and novel %) across a range of depths in the quadruple-sequenced Qatari, and comparing the before/after filtering total variants and novel SNP rates both across platforms and with published reports of a similar analysis[17]. After application of batch-specific filters, the SNP data from the three platforms was integrated using GATK. Using the calls of sites covered in the three batches for the n=1376 individuals, population structure analysis was conducted in combination with 1000 Genomes Phase 3[18] and the Human Origins dataset[19] using ADMIXTURE[20]. Each individual was assigned to one of 12 ancestral population clusters based on their dominant ancestry, and the validity of the clustering was confirmed using Principal Components Analysis[21]. Relatedness analysis was conducted using KING[22], and first and second degree relatives assessed using a liberal cutoff (to assure an unrelated sample) were excluded. Using the remaining 1,005 unrelated Qataris, including 917 exomes and 88 genomes, allele frequencies at SNPs was calculated. In addition, indels were called in the 88 genomes using the CASAVA[23] pipeline and an assessment of size and allele frequency of these variants was also conducted.

In order to facilitate genome interpretation for precision medicine in Qatar, major allele SNPs and indels were identified, and the GRCh37 reference genome was modified at these sites to produce the Qatar Reference Genome (QTRG). At each site in the QTRG reference, the major allele in Qatar is the reference allele in the sample of n=1005 unrelated Qataris. Three versions of the QTRG were produced, including versions incorporating major allele SNPs (QTRG1), major allele indels (QTRG2), and major allele SNPs and indels (QTRG3).

In order to select an optimal reference for analysis of n+1 genomes and exomes, the three references were compared in terms of mapped read depth. Genome sequence data from the quadruple-sequenced Qatari were mapped to the four references (GRCh37, QTRG1, QTRG2, and QTRG3) using BWA[24], and the depth of coverage was calculated using GATK for all sites and for modified sites. The resulting depth was compared across platforms, and the reference with the deepest resulting coverage was selected for further analysis. Further inspection of the value of the reference for variant detection was assessed by producing variant calls using GATK Best Practices twice, with the only change being the reference genome used. This comparison was conducted for the quadruple-sequenced Qatari, a Qatari family of n=15 genomes of Persian ancestry, and a diverse panel of n=16 Qatari exomes. The number of variants identified was com-pared across references. In order to assess the impact beyond modified variants, the expected reduction in variants (based on the number of modified variants in the individual) and the observed reduction in variants was compared.

An obstacle to using a QTRG reference containing common indels is that the genomic coordinates of known variants and genes are shifted. Conversion of coordinates between references is conducted using a "liftover", where the GRCh37 and QTRG chromosomes are aligned using progressiveCactus[25], and a liftover for known gene and variant positions from GRCh37 to QTRG is conducted using HalTools[26]. This conversion was conducted for variants functionally annotated to have a known link to disease and are high priority for future studies of Mendelian disease in Qatar. Through a combination of automated and manual curation following the most recent ACMG guidelines for next-generation sequencing interpretation[10], liftover annotation of n=128 pathogenic variants was conducted. For further methodologic details, see Supplemental Methods.

## Results

**Data Integration**

In order to build a reference genome for precision medicine applications in Qatar and the greater MENA region where major allele variants are incorporated into the reference sequence, n=1161 Qatari and n=215 non-Qatari living in Qatar were sequenced on the Illumina platform in three batches (n=108 genome, n=67 Exome37Mb and n=1201 Exome51Mb) to 38* genome depth and 70* exome depth. Genotypes were generated for each batch using the GATK Best Practices workflow[16,27], and combined into a single variant call set after application of batch-specific filters (see Supplemental Figure 1 for workflow overview). In the integrated call set, the novel SNP rate was assessed for genomic intervals covered in all samples, a per-sample mean of 0.73% in the genome batch and 0.72% in both exome batches (Supplemental Table I). The plat-form specific filters evaluated included minimum depth (ranging from 2* to 20*) and minimum allele count (ranging from 1 to 6). Increasing minimum depth, but not increasing minimum allele count had an effect of increasing the median depth of exome sequencing (Supplemental Figure 2A), and decreasing the total number of variant sites observed in exome sequencing (Supplemental Figure 2B), while neither filter had major impact on the genome sequencing median depth nor total variant sites. A minimum depth and minimum allele count was applied to the two batches of exome sequence data (Exome38Mb, n=67 and Exome51Mb, n=1201) such that the novel SNP rate was the same or lower than the genome rate for 12* minimum depth and minimum allele count 1. At this threshold, the Exome38Mb filter was minimum 12* depth (minimum allele count 1), and the Exome51Mb filter was minimum 10* depth (minimum allele count 2) (black line in Supplemental Figure 2C).

Using a quadruple-sequenced female Qatari as a benchmark, the number of batch-specific variants was assessed for all variants and for novel variants before and after application of the batch-specific filters (Supplemental Figure 3). All four datasets were analyzed using the same pipeline, and a comparison was conducted on CCDS coding intervals covered in all four platforms (within the Exome38Mb target intervals). Before filtering, the rate of batch-specific variants in the full call set was 10.1% (Supplemental Figure 3A), and an excess of batch-specific novel variants (82.6%, Supplemental Figure 3B) was observed. After filtering, the batch-specific variant rate was reduced to 4.9% (Supplemental figure 3C), and the rate of batch-specific novel variants was considerably reduced (16.8%; Supplemental Figure 3D).

The impact of the filters on variant sensitivity and novel SNP rate was assessed across a range of mean depth for the four platforms (Supplemental Figure 4). Variant sensitivity increases with additional depth, however sensitivity reaches a plateau after 25* depth for genome sequencing and 65* depth for exome sequencing (Supplemental Figure 4A). Application of batch-specific filters reduced the sensitivity of exome sequencing to a greater extent than genome sequencing (Supplemental Figure 4B). In terms of novel SNPs, a linear increase in novel variant rate was observed with increasing depth across platforms, with the exception of the HiSeq 2500 genome, which reaches a plateau under 1% (Supplemental Figure 4C). After filtering of the Exome38Mb and Exome51Mb data, a similar plateau effect was observed for exome data (Supplemental Figure 4D). The filters were verified to not be overly stringent, based on the total variants per genome or exome in coding regions being in the range of prior studies[17].

**Ancestry Analysis**

Prior studies of the Qatari population have characterized it as a diverse population with influences of Arab, Bedouin, Persian, South Asian, and African ancestry[28]. In order to characterize the ancestry of our sample, the n=1161 Qataris (n=108 genomes, n=1053 exomes) were compared to n=215 non-Qataris living in Qatar and sampled by exome sequencing in this study, as well as public databases of diverse genomes, including the 1000 Genomes Project[18,29] and the Human Origins data (described in Supplemental Table II). Using the parameter K=12 (12 ancestral populations) that was inferred to be optimal in a prior study[28], the ancestral population structure of the combined sample of n=5661 genomes was analyzed using ADMIXTURE on a set of n=2265 SNPs segregating in all four datasets (Qataris, non-Qataris sampled in Qatar, 1000 Genomes, Human Origins). The proportion of 12 ancestries was determined for each individual, and individuals were assigned to a cluster based on the dominant ancestral population in their genome (Supplemental Figure 5A and Supplemental Table III). The Qataris were assigned to 7 clusters, determined to be of European (K=1, n=5 Qataris), South Asian (K=4, n=82 Qataris), Bedouin (K=5, n=566 Qataris), African Pygmy (K=6, N=1 Qatari), Bedouin (K=8, n=236 Qataris), Persian (K=9, N=194 Qataris), and Sub-Saharan Africa (K=10, n=77 Qataris) (Supplemental Table III and Supplemental Figure 5B). Using a color-coding scheme based on the 12 ancestral clusters (Supplemental Table III), a principal components (PC) analysis was conducted for the combined set of n=5661 samples. Clearly separation of African and non-African clusters were observed when plotting PC1 *vs* PC2 (Supplemental Figure 6A), and resolution of European, Asian and Middle Eastern clusters was observed when plotting PC2 *vs* PC3 (Supplemental Figure 6B).

**Relatedness Analysis**

The Qataris and non-Qataris sampled in this study included a mixture samples from studies of rare and common diseases, including both families affected with Mendelian disorders and a randomly sampled group of presumably unrelated type-2 diabetics and controls. Given the within-family sampling and the known high rate of consanguineous marriage in Qatar[30], we sought to exclude relatives prior to estimation of variant allele frequency in the general Qatari population. For this purpose, an analysis of relatedness was conducted on the n=1376 individuals sequenced in this study, using SNP variants in genomic intervals covered in the intersection of the three batches (within Exome38Mb target intervals). Using an LD-pruned set of n=381028 SNPs, the relatedness was calculated across all pairs of individuals. A total of n=736 relationships were observed, including n=239 1st degree relationships, n=71 2nd degree relationships, and n=526 3rd degree relationships (Supplemental Table IV). The relationships were plotted using Cytoscape[31], color-coded by inferred ancestry, and the majority of relationships were between individuals of the same ancestry. The largest pedigrees recovered were of Middle Eastern (Bedouin, Arab, Persian) ancestry, confirming theories of deep population structure among Qataris and within-tribe intermarriage (Supplemental Figure 7). After exclusion of 1st degree and 2nd degree relatives, a total of n=1,005 Qataris remained in the analysis, including n=88 genomes and n=917 exomes (n=64 Exome38Mb and n=853 Exome51Mb).

**Variant Discovery in Qatar**

The individual variants observed in 1005 Qataris were aggregated and their allele frequency was quantified. After exclusion of relatives and application of batch-specific filters, an average of 4,045,064 SNPs were observed per genome (n=88), an average of 15,382 SNPs were observed per Exome51Mb (n=853) and an average of 13,538 SNPs were observed per Exome38Mb (n=64). The novel SNP rate was higher in the genome (1.99%) than in the exome, with a slightly higher rate in the Exome51Mb (0.99%) than in the Exome38Mb (0.67%) samples. This trend is consistent with higher degree of selection on protein coding genes as compared to non-coding DNA, and the inclusion of transcripts that do not code for protein in the Exome51Mb targets ( ). The rate of novel variants was higher in ChrX, ChrY, and MtDNA for the genome data, possibly due to a bias towards autosomal data in dbSNP (Table I). The SNP data was aggregated across platforms, resulting in a total of 20,937,965 SNPs, of which 14.21% were not previously observed in dbSNP (as of build 146) (Supplemental Table I). A call set of short indel variants (< 300bp) was generated for the 88 unrelated Qatari genomes using the CASAVA pipeline, identifying a total of 5,452,613 variants, of which 58.37% were novel.

**Construction of the Qatar Reference Genome**

The allele frequency was quantified for each SNP, and a total of 1,931,122 (9.22%) of the SNPs were present in over half the Qatari alleles sampled, and hence are candidates for modification in the Qatar Reference Genome (Supplemental Table V). Furthermore, while prior studies of ancestry-specific reference genomes have incorporated major allele SNP variants, the inclusion of major alternate allele (MAA) indel variants has not previously been explored. In order to explore the value of incorporating indels into the reference, 1,882,405 MAA indel variants (34.52%) were identified (Supplemental Table V).

Three versions of the Qatar Reference Genome were constructed, where major alternate alleles (MAAs) were modified. The first version (QTRG1) includes modification of MAA SNPs, the second version (QTRG2) includes modification of MAA indels, and the third version (QTRG3) includes modification of MAA SNPs and indels.

**Selection of an Optimal Reference for Read Mapping**

In order to determine which reference produces the greatest improvement in terms of mapped read depth, the genome of the quadruple-sequenced Qatari was mapped to the four reference genomes, including the unmodified reference (GRCh37), and the Qatari references including modifications at MAA SNPs (QTRG1), indels (QTRG2), and both SNPs and indels (QTRG3). An improvement in mapped read depth was observed for all Qatari references, with the greatest improvement (10%) at modified sites when using QTRG3 (Figure 1A). A modest improvement in mapped read depth was observed overall, with the benefits extending beyond modified sites (Figure 1B).

**Impact of Using Qatar Reference Genome on Variant Sensitivity**

Increased mapped read depth results in increased sensitivity for variant detection (Supplemental Figure 4A). In order to assess the benefit of improved mapped read depth using QTRG3 on variant sensitivity, GATK Best Practices analysis was conducted twice, with the only difference between iterations being the reference used (GRCh37 versus QTRG3). This analysis was conducted for the quadruple-sequenced Qatari, for a 15-member family of Qataris of K=9 Persian ancestry, and a diverse panel of n=16 Qatari exomes.

In order to quantify the impact of using GRCh37 versus QTRG3 as a reference on the quality of variant genotypes, a single Qatari was quadruple-sequenced (2 exomes and 2 genomes) and mapped to both references. After the analysis was completed, a total of n=8 callsets were produced and compared. The focus of the analysis was sites covered with at least 12* depth in all four platforms. Within these intervals, no discordant genotypes were observed within the four GRCh37 calls nor within the QTRG3 calls. The coverage depth was 1* to 2* higher at variant sites in the genomes, while coverage depth was 4* to 6* lower at variant sites in the exomes (Supplemental Table VI). A 10.3% reduction of variants was observed, lower than expected given that 42.28% of the GRCh37 variants are MAAs incorporated into the reference.

In order to quantify the impact of using GRCh37 versus QTRG3 on sensitivity for variants beyond those modified in the reference, Illumina paired-end 100 bp genome sequencing reads for n=15 Qataris from a family of Persian ancestry were mapped to both GRCh37 and QTRG3. An average of 23% (7*) depth improvement was observed when using QTRG3 (Supplemental Table VII). The number of variants observed per genome was reduced on average n=756,671 (16%), however this was on average 25% lower than expected, based on an average of 41% modified sites in each genome (Supplemental Table VII).

Given the diversity of the Qatari population, use of the QTRG reference may not provide the same benefit for all ancestries. In order to quantify ancestry-specific differences in depth and variant sensitivity, genome analysis using both GRCh37 and QTRG3 was conducted for a diverse panel of n=16 Qatari exomes. In contrast to the genome, across all ancestries a reduction of variants (up to 76%) was observed to be in excess of the expected (up to 42%, based on modified sites). A significant difference in the reduction was observed between Qataris of Sub-Saharan African ancestry and Qataris of Bedouin or Arab ancestry (one-tailed t-test p value < 0.01) (Supplemental Table VIII).

**Liftover of Mendelian Disease Variants in Qataris**

A major challenge for use of the QTRG3 reference that incorporates MAA SNPs and indels is the migration of variant positions due to indels. The position of known variants and of genes is different in QTRG3 than in GRCh37. Hence, a major obstacle to use of QTRG3 in precision medicine studies is the lack of genome interpretation databases on QTRG3 coordinates. A similar issue arises in genomics when a novel assembly of the human reference genome is produced, the translation of coordinates from one assembly to another is known as a "liftover"[26]. There is an established methods to accomplish this task, involving pairwise sequence alignment of pairs of chromosomes, such as GRCh37 Chr20 and QTRG3 Chr20, and then using the alignment the coordinates for a list of sites on GRCh37 Chr20 are "lifted over" to QTRG3 Chr20 coordinates.

Variants were annotated using SNPEFF[32] (Supplemental Table IX.), and potentially deleterious SNPs in coding genes were then grouped into three categories. Out of n=20,864,277 SNPs, a total of n=155,571 are potentially pathogenic protein coding SNPs (Category 3), including n=50,757 in genes linked to a phenotype (Category 2) and n=2,152 of these that are variants with known links to a phenotype (Category 1; Table II). Based on ACMG recommendations[10], only the n=1,020 (999 + 21) Category 1 at minor allele frequency below 5% (rare) are retained for further consideration (Table II). The rare Category 1 variants were further filtered to exclude n=200 singletons (single alleles in the population), which are enriched for false positives (Supplemental Figure 3). The literature for the remaining variants was reviewed, and n=122 variants linked to a phenotype that can clearly be defined as a Mendelian (dominant, recessive, x-linked) disease are presented in Supplemental Table X. By using the liftover method, the QTRG3 coordinates for these pathogenic variants was ascertained (Supplemental Table X).

## Discussion

This study presents a set of publicly available bioinformatics tools and resources for genome interpretation studies in Qatar and closely related MENA populations. Over 1000 Qataris were sequenced to produce this resource, effectively sampling nearly 0.4% of the indigenous population of Qatar. Of 26 million SNPs and indels observed in the autosomes, sex chromosomes, and mitochondrial DNA, over 9% were in fact the major allele in Qatar. Using the com-plete catalog of variants, a reference genome custom tailored to disease research in the Qatari population was constructed, named here the Qatari Reference Genome (QTRG).

The value of utilizing the QTRG *vis-a-vis* the standard reference genome GRCh37 was demonstrated through improved read depth and variant sensitivity. Use of this reference in Qatar will lead to higher quality interpretation for both individual genomes and Mendelian disease studies. In order to facilitate Mendelian disease research using the QTRG, a catalog of known pathogenic mutations in Qatar was compiled, with genomic coordinates on both the GRCh37 and QTRG included.

Although prior studies have constructed reference genomes tailored to distinct ancestries[8], this is the first such study that incorporates both SNPs and indels into the reference, and solves crisis of genome interpretation that would ensue without liftover of genomic coordinates for known variants. While in the near future *de novo* assembly of personal genomes could become routine[33], the reference genome is expected to remain useful for comparisons across a large sample of genomes, and for comparison to public databases. As expressed by the version information for QTRG (version 1, 2 and 3), future versions of the Qatari Reference Genome are therefore planned for release, based on inclusion of major alleles for a broader spectrum of genetic variation (such as CNVs, genome rearrangements, and short tandem repeats), and ongoing sampling and sequencing of a larger representation both within Qatar and in the MENA region.

**Acknowledgments**. We thank 1000 Genomes Project collaborators, as well as colleagues in the American Society of Human Genetics and the Society for Molecular Biology and Evolution for valuable discussion on genome analysis methods; N. Mohamed for editorial support.

**Conflict of Interest:** None, for all authors.

**Data Access:** Genome and submitted to the Short Read Archive (SRA) section of the NCBI SRA database (SRA accession SRP061943). The Qatar Reference Genome (QTRG) sequence, the database of annotated variants in Qatar, and bioinformatics tools for analysis of genomes using QTRG are available at our website (http://geneticmedicine.weill.cornell.edu/genome).

**Table I. Individual Variant Discovery in 1005 Unrelated Qatari[1]**

| Statistic | Genome (n=88) | | | | | 51 Mb Exome (n=853) | | | | 38 Mb Exome (n=64) | | | |
|---|---|---|---|---|---|---|---|---|---|---|---|---|---|
| | Individual | Autosomes | ChrX | ChrY | MtDNA | Individual | Autosomes | ChrX | ChrY | Individual | Autosomes | ChrX | ChrY |
| Variant sites | 4,045,064 | 3,893,076 | 151,017 | 937 | 34 | 15,382 | 15,069 | 312 | 1 | 13,839 | 13,538 | 300 | 1 |
| Novel variant sites | 96,774 | 77,366 | 19,359 | 42 | 7 | 159 | 152 | 7 | 0 | 96 | 91 | 5 | 0 |
| Novel variant rate | 2.39% | 1.99% | 12.82% | 4.48% | 20.59% | 1.03% | 0.99% | 2.24% | 0.00% | 0.69% | 0.67% | 1.67% | 0.00% |
| Alternate alleles | 5,510,301 | 5,311,794 | 196,565 | 1,874 | 68 | 21,046 | 20,613 | 431 | 2 | 19,116 | 18,709 | 405 | 2 |
| Novel alternate alleles | 98,792 | 78,916 | 19,778 | 84 | 14 | 160 | 153 | 7 | 0 | 97 | 92 | 5 | 0 |
| Novel allele rate | 1.79% | 1.49% | 10.06% | 4.48% | 20.59% | 0.76% | 0.74% | 1.62% | 0.00% | 0.50% | 0.49% | 1.23% | 0.00% |
| Heterozygous sites | 2,594,268 | 2,489,084 | 105,184 | - | - | 9,759 | 9,564 | 195 | - | 8,897 | 8,698 | 199 | - |
| Novel heterozygous sites | 94,762 | 75,772 | 18,990 | - | - | 157 | 150 | 7 | - | 94 | 89 | 5 | - |
| Novel heterozygous rate | 3.65% | 3.04% | 18.05% | - | - | 1.61% | 1.57% | 3.59% | - | 1.06% | 1.02% | 2.51% | - |
| Mean depth at variant site | 41 | 41 | 40 | 20 | 250 | 63 | 64 | 69 | 42 | 64 | 64 | 74 | 27 |
| Mean depth at novel variant site | 41 | 41 | 39 | 20 | 250 | 59 | 60 | 60 | 47 | 63 | 63 | 76 | - |
| Transition:transversion ratio | 2.03 | 2.03 | 1.78 | 1.51 | 33.00 | 3.18 | 3.18 | 2.77 | - | 3.25 | 3.26 | 2.67 | - |
| Novel transition:transversion ratio | 1.33 | 1.35 | 1.09 | 1.63 | - | 0.77 | 0.77 | 0.75 | - | 1.58 | 1.56 | 1.50 | - |

[1] Shown is a summary of the average number of variants observed per individual, identified in 917 unrelated Qatari exomes and 88 unrelated Qatari genomes. Variants were genotyped separately for autosomes, X in males, X in females, Y in males, and mtDNA. 99.8% of X variants in males were also observed in females, hence summary statistics are based on female chromosomes. Shown is the average per individual of number of variant sites, number of variant alleles, the transition-to-transversion ratio (Ts:Tv) of variants, and the % not in dbSNP (novel).

Table II. Variants in Qatar, Stratified by Allele Frequency and Potential for Pathogenicity[1]

| | All variant alleles (GRCh37) | | Major reference allele (MRA) | | | | | | Major alternate allele (MAA) | | | | | |
|---|---|---|---|---|---|---|---|---|---|---|---|---|---|---|
| | | | Rare alternate allele (< 5% alternate allele frequency) | | Common alternate allele (5% to 50% alternate allele frequency) | | Common reference allele (50% to 95% alternate allele frequency) | | Rare reference allele (95% to 100% alternate allele frequency) | | Unobserved reference allele (100% alternate allele frequency) | |
| Category | n | % | n | % | n | % | n | % | n | % | n | % |
| All SNPs | 20,864,277 | 100.00 | 12,948,368 | 62.06 | 5,938,490 | 28.46 | 1,693,649 | 8.12 | 195,466 | 0.94 | 88,303 | 0.42 |
| 3- Potentially pathogenic | 155,571 | 0.75 | 124,947 | 0.60 | 23,282 | 0.11 | 5,957 | 0.03 | 956 | <0.01 | 428 | <0.01 |
| 2 - In gene linked to phenotype | 50,757 | 0.24 | 40,894 | 0.20 | 7,445 | 0.04 | 2,002 | 0.01 | 290 | <0.01 | 125 | <0.01 |
| 1 - Variant with known link | 2,152 | 0.01 | 999 | <0.01 | 876 | <0.01 | 253 | <0.01 | 21 | <0.01 | 2 | <0.01 |

[1] The major allele variants are modified in the QTRG genome, such that all reported variants are the minor allele. A total of 230,395 potentially deleterious SNPs in the 917 exomes and 88 genomes were computationally categorized with respect to allele frequency and databases of genes and variants with reported links to a phenotype. Variants were assigned to genes and their function was predicted with respect to ENSEMBL[34] gene models using SNPEFF[32], and potentially deleterious coding SNPs (nonsynonymous, splice donor site, splice acceptor site, stop gain, start loss) variants were extracted for further analysis. A database that combines OMIM[35], HGMD[36], GWAS[37], PharmGKB[38], Human Phenotype Ontology[39], and ClinVar[40] was compiled, where these annotations were used to divide the potentially deleterious variants into three categories, variant and gene linked to a phenotype (Category 1), gene but not variant linked to a phenotype (Category 2), and neither variant nor gene linked to a phenotype (Category 3). The totals for each category are shown in the left-most columns, including number and percentage. These variants were then sub-classified into two major (major reference allele, major alternate allele) and 5 minor categories based on variant allele frequency in Qatar rare alternate allele (up to 5% variant allele frequency), common alternate allele (between 5% and 50% allele frequency), common reference allele (from 50% to 95% alternate allele frequency), rare reference allele (from 95% to 100% alternate allele frequency), unobserved reference allele (100% allele frequency). The major alternate alleles (MAA) are modified in the Qatar Reference Genome (QTRG).

**Figure Legends**

**Figure 1.** Differences in mapped read depth across reference genomes. In order to select the optimal reference for analysis of Qatari genomes and exomes, the mapped read depth was compared between GRCh37 and three alternative reference genomes based on major alternate alleles (MAA) observed in n=1005 Qatari. Illumina paired-end 100 bp reads for 37* genome sequencing of a female Qatari were mapped using BWA to GRCh37, QTRG1, QTRG2, and QTRG3 reference genomes. The differences between the three Qatari references is that QTRG1 incorporates MAA SNPs, QTRG2 incorporates MAA indels, and QTRG3 incorporates both MAA SNPs and MAA indels. The depth of coverage was measured at **A.** across the genome and **B.** at MAA sites modified in the QTRG.

## A. Modified sites

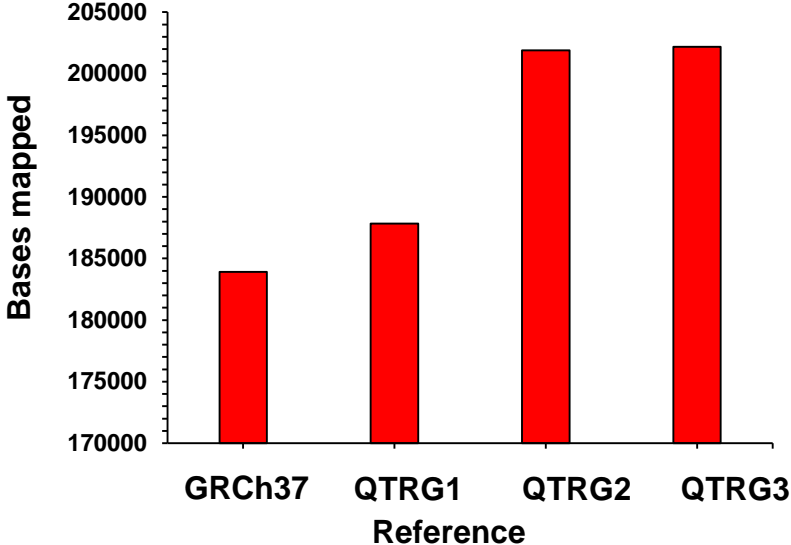

## B. All sites

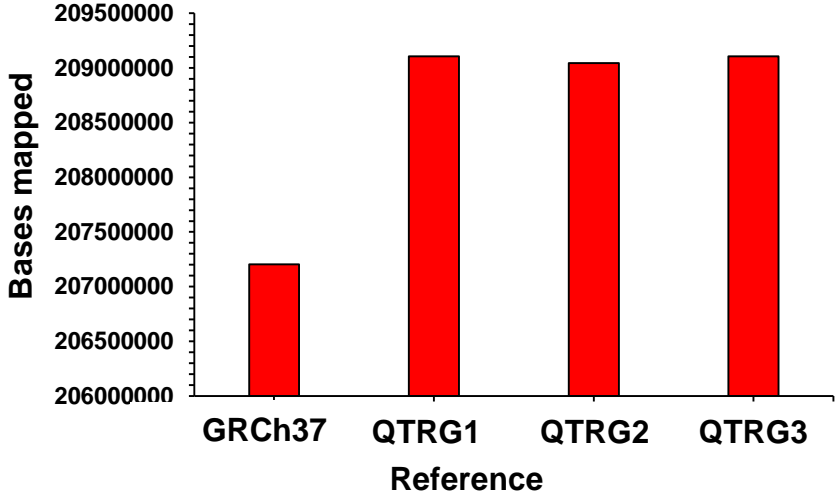

**Supplemental Methods**

**Inclusion Criteria**

In order to maximize the accuracy of Qatari ancestry inference, both Qataris and non-Qatari Non-Qataris living in Qatar were selected for sequencing. As selection criteria for Qataris, subjects were required to be third generation Qataris where all four grandparents were Qatari citizens born in Qatar, as assessed by questionnaires. Recent immigrants or residents of Qatar who traced their recent ancestry to other geographic regions were classified as Non-Qataris. DNA was extracted from blood using the QIAamp DNA Blood Maxi Kit (Qiagen Sciences Inc., Germantown, MD).

DNA samples were collected for a total of 1,376 unique individuals. Samples included in this study were obtained from two separate studies, the first being a study of diabetes in unrelated Qataris (1,084), and the second being a study of Mendelian disorders in Qatari and Non-Qatari families living in Qatar (292).

**Ancestry Informative Marker Panel Genotyping**

An initial estimate of the ancestry of each Qatari was conducted on all 3-generation Qataris. Based on prior studies of Affymetrix 5.0[1,2] and exome sequencing[1,3,4] of Qataris, most Qataris can be placed into one of three ancestry subpopulations: Bedouin (Q1), Persian/South Asian (Q2) and African (Q3)[1,4]. An estimate of the relative proportions of tripartite ancestry was also estimated in each Qatari based on genotypes for a panel of 48 ancestry-informative SNPs[4] ascertained using the TaqMan platform (Life Technologies, Carlsbad, CA), followed by STRUCTURE[5] analysis with k=3, where individuals with at least 65% ancestry in one of three clusters were classified as "low admixture" Qataris. Qataris were recruited for genetic studies at Hamad Medical Corporation clinics, and consenting volunteers were initially genotyped on the 48-SNP ancestry panel. Individuals with a high proportion of ancestry in one of three clusters were prioritized for sequencing. DNA was extracted from blood using the QIAamp DNA Blood Maxi Kit (Qiagen Sciences, Germantown, MD).

**Genome Sequencing**

In order to characterize the spectrum of Qatari genetic variation, 108 Qataris were sequenced to a median depth of 37x on the Illumina platform. Next-generation sequencing libraries were generated using the Illumina TruSeq protocol, and sequencing was conducted at the Illumina FastTrack Services sequencing facility using the HiSeq 2500[6]. Sufficient paired-end 100 bp reads were generated in order to produce a median of 112 GB of sequence data passing filters and aligned to the hg19 human reference genome with a median insert size of 301 bp, where at least 85% of bases ≥Q30, passed filtering steps and were aligned. Among non-N bases in the reference genome, at least 98% were covered by at least one base in all 108 genomes. Reads were realigned to the 1000 Genomes Project version of the GRCh37 human reference genome using BWA 0.5.9[7] (maximum insert size 3 kb), and mapped reads were prepared for variant calling using GATK Best Practices[8], including PCR duplicate removal using SAMtools[9], producing an average of 37x depth in autosomal chromosomes, with a mean of 98% of mappable sites covered per genome. SNP genotypes for all 108 genomes were simultaneously called using GATK[10]. In addition, short indel (<300 bp) calls were produced using the CASAVA 1.9 pipeline by Illumina.

**Exome Sequencing**

Exome sequence data was produced for 1,268 Qataris. In addition, 53 technical replicates were produced, a total of 1,321 exomes sequenced and analyzed, where the entire sample included 1,053 unique Qataris and 215 unique Non-Qataris. A total of 1,221 exomes were sequenced by the New York Genome Center (NYGC), an additional 100 Qatari were sequenced at Beijing Genomics Institute (BGI)[4]. The NYGC exomes were produced using Agilent SureSelect Human All Exome 51 Mb targets, and are hearafter referred to as "Exome51Mb", while the BGI exomes were produced using Agilent SureSelect Human All Exome 38 Mb targets, hereafter referred to as "Exome38Mb".

The Exome51Mb batch was produced using standard methods. Exome DNA enrichment was performed using Agilent SureSelect Human All Exon Kit V5 (Agilent Technologies, Santa Clara, CA) hybrid capture targets for 51 Mb of coding regions, with sequencing on the Illumina HiSeq 2500 platform[6]. Reads were mapped to GRCh37 using BWA v0.5.9[7], mean coverage depth was verified to be greater than 30x for all autosomal chromosomes in all samples. Quality filtering of mapped reads was conducted using the GATK Best Practices workflow[8], including realignment across known indels, recalibration of base quality scores, and removal of duplicate reads.

**Technical Replicates**

Technical replicates in the study included 5 sequenced in triplicate (genome, Exome38Mb, Exome51Mb) and 43 that were sequenced on two platforms, including 26 genome and Exome38Mb (31 total, including the 5 triplicates), 4 sequenced genomes Exome51Mb (9 total, including the 5 triplicates), 2 sequenced in both Exome51Mb and Exome38Mb batches (7 total), and 11 sequenced twice in the Exome51Mb batch. The genotypes for these technical replicates were compared to assess overall reproducibility across sequencing platforms. The genome of the triplicate (a Qatari female) was subsequently sequenced on a fourth platform (Illumina HiSeq X), in order to have one quadruple-sequenced Qatari to use as a benchmark for analysis.

**Integrated Analysis of Exome and Genome Samples**

In order to conduct analysis of population structure and relatedness on the maximum number of samples, genotype data from the 108 genomes was integrated with exome sequence data for 1,268 Qataris. Including the technical replicates, a call set for all 1,429 samples was produced by simultaneous genotyping using the GATK UnifiedGenotyper algorithm[10]. Genotype-ready BAM alignment files were prepared as described above for genome, Exome51Mb, and Exome38Mb samples. Genotype calling was limited to sites in the Agilent SureSelect 38Mb targets that overlap CCDS build 15 coding exons[11]. Variants observed in at least one sample with Q30 or higher were reported. All bi-allelic SNPs genotyped successfully in at least 80% of samples were kept for further analysis. In order to facilitate downstream analysis, variant VCF files were converted to PLINK[12] format using PLINK2[13].

**Optimization of Batch-Specific Genotype Filters**

Integration of data from three distinct batches of sequence data produced at three distinct sequencing centers on three distinct sequencing library preparation kits and three different versions of the Illumina sequencer introduces a number of variables that could potentially result in-batch effects in the variant calls. Novel variants, defined as those not present in dbSNP (build 146 being the latest at the time of writing), singletons (single alleles observed in only one individual in the population sample), and genotypes with low sequencing depth are typically enriched for sequencing errors and batch artifacts[14]. Hence, one way to control for batch effects is to filter variants below a threshold depth $d$ and threshold variant allele count $v$, such that the mean novel SNP rate is consistent across batches within genomic intervals covered by the intersection of all batches. In order to optimize $d$ and $v$, two approaches were followed. First, the novel SNP rate for a range of $d$ and $v$ thresholds were calculated in each batch, and batch-specific $d$ and $v$ that produced the highest possible consistent novel SNP rate was selected and applied to each call set. Second, the impact of the selected batch-specific filters $d$ and $v$ were applied to the quadruple-sequenced Qatari, and the proportion of batch-specific variants (defined as variants observed in only one of the four batches in the quadruple-sequenced Qatari) was compared before and after application of filters. The proportion of batch-specific variants was calcu-lated for both all variants and novel variants, with the aim of controlling the batch-specific variants at below 5%.

A major concern when applying genotype filters is the risk of reducing sensitivity for known and novel variants. In order to confirm that overly-stringent filtering was applied, the impact of the selected filters on sensitivity and novel SNP rate was assessed across a range of depths for the quadruple-sequenced Qatari, in order to ensure that the number of variants observed within CCDS coding exons (within the Exome38Mb target intervals) was in the range of published benchmarks of per-exome variant counts[15], and that excess novel SNP rates were avoided.

**Population Structure**

In order to quantify admixture in 1,161 Qataris and 215 Non-Qataris living in Qatar, the integrated SNP genotypes of 1376 individuals within Agilent SureSelect 38 Mb targets were integrated with 4,235 genomes from public databases. The public databases included 1,862 from Human Origins (HumOr), 1,028 from 1000 Genomes Phase 1 not in HumOr (1000Gp1), and 1,345 from 1000 Genomes Phase 3 not in HumOr nor 1000Gp1 (1000Gp3). The integrated variant dataset included 5,611 SNPs segregating in all datasets (Qatar Genomes, Qatar Exomes, HumOr, 1000Gp1, 1000Gp3) after filtering to exclude SNPs with more than 20% missing genotypes, SNPs with Hardy Weinberg equilibrium p value $< 1 \times 10^{-6}$ (as calculated using PLINK2), and SNPs removed by linkage disequilibrium pruning (conducted using the PLINK2 –indep-pairwise function in a sliding window of 1000 SNPs, step size 25 SNPs, and exclusion of SNPs with $r^2 > 0.1$). Inference of ancestry was conducted using ADMIXTURE[16] for the 5,611 individuals. Ancestry proportion inference was conducted using K=12, based on lowest cross-validation error in a prior study[17]. For each individual, ADMIXTURE outputs the proportion of ancestry in each ancestral population (K). Each individual was assigned to the ancestral population with the largest proportion in his or her genome.

**Relatedness**

Given the high rate of consanguineous marriage in Qatar[18], and the inclusion of complete families in our study, the potential over-sampling of relatives and families in sample of the population is high. In order to avoid over-sampling or alleles from relatives, and produce allele frequency estimates for the general Qatari population, the pairwise relatedness between 1,376 samples (including Qataris and Non-Qataris living in Qatar, after exclusion of technical replicates) was calculated using KING[19], and 1st/2nd degree relatives were excluded. Genotypes used for comparison were limited to CCDS coding intervals covered in the intersection of the three sequencing batches (n=108 genome, n=67 Exome38Mb, and n=1201 Exome51Mb). Given the high level of inbreeding in the Qatari population[18], and the use of coding variant data for the analysis, the pairwise relatedness estimates of KING and related software are expected to over-estimate relatedness[19-21], such that assessment of 1st/2nd degree relatedness is expected to be a liberal and therefore over-assessment of relatedness in the sample. Using this liberal cutoff, a total of 1005 unrelated Qataris (n=88 genome, n=64 Exome38Mb, and n=853 Exome51Mb) were identified, resulting in the exclusion of 156 1st/2nd degree Qatari relatives.

**Construction of the Qatari Reference Genome (QTRG)**

Total variants were counted in each chromosome of the n=88 unrelated Qatari genomes and n=917 unrelated Qatari exomes, and an aggregate variant list was compiled. Variant allele frequency was calculated using PLINK2[12] and VCFtools[22]. Novel variants (defined as not present in dbSNP build 146) were quantified in both individual and aggregate sets, as was the Ts:tv ratio (measured using VCFtools[22]). Variants in Qatar were divided into 2 categories based on which allele is the major allele, subsequently described as major reference alleles (MRA) and major alternate alleles (MAA)

Three versions of the Qatar Reference Genome were produced. In order to produce an initial version of the Qatari Reference Genome (referred to as "QTRG1"), the standard reference genome (GRCh37) was modified at all sites where the variant allele in Qatar is the major allele in Qatar, in this case including all MAA SNPs. The modifications to the GRCh37 FASTA[23] file was conducted using VCFtools[22], which takes the reference genome and a VCF file with the major allele variants as inputs, and outputs a modified FASTA file. QTRG1 was prepared for use in analysis of human genomes and exomes by creation of indexed versions of the reference using SAMtools[9], Picard[7,24], and BWA[7]. A second version of the Qatar Reference Genome (QTRG2) was produced by including all MAA indels. A third version of the Qatar Reference Genome (QTRG3) was produced by including the MAA SNPs and MAA indels.

**Assessment of QTRG *vs* GRCh37 on "n+1" Qatari**

The value of the QTR1 population-specific reference genome for assessing major and minor allele variants was demonstrated by genome interpretation of the next Qatari sequenced, referred to the "n+1" genome, defined as the "next" genome sequenced and analyzed using data from the panel of n=1005 genomes analyzed thus far. The n+1 analysis was conducted by taking the sequence data for an individual genome, and processing it twice using the same GATK Best Practices Workflow[8], with the only difference being the reference genome used. Three experiments were conducted in order to assess the value of using QTRG. First, genome sequence data for the quadruple-sequenced Qatari was mapped to four references (including GRCh37, QTRG1, QTRG2, and QTRG3), and differences in mapped read depth at both modified and unmodified sites was compared. The reference with the greatest mapped read depth (in this case QTRG3) was selected for the two other experiments. Second, a family of 15 genomes sequenced at 37x depth using paired-end Illumina 100bp sequencing was analyzed using the GATK Best Practices Workflow[8] twice, with the only change being the reference genome used (GRCh37 or QTRG3).

The depth and number of variants was compared between the two references for each individual. In addition, in order to assess the impact of using the reference beyond modified sites, the observed reduction in number of variants was compared to the expected number, where the expected reduction is the number of number of "flipped" variants (variants incorporated into the QTRG3 reference) present in the individual genome. The difference across the 15 genomes was tested for significance using a 2-df Chi-squared test. Third, the analysis was repeated on a diverse panel of n=16 Qatari exomes, and differences in the increased depth and reduced variant count was compared across populations with a one-tailed t-test.

**Gene and Variant Annotation Liftover**

A major issue for utilizing the QTRG3 reference genome in n+1 genome interpretation is the need to update the positions of known genes and variants. In genomics this is known as a "liftover", an analysis that is conducted each time a new version of the human reference genome assembly is produced. In order to assess the feasibility QTRG3 for genome interpretation, a liftover analysis was conducted for known pathogenic variants observed in filtered call set for the n=1005 unrelated Qatari. The variants with respect to GRCh37 were annotated, and then the variant coordinates were "lifted over" to QTRG3 coordinates. The liftover process has two steps, first each chromosome of GRCh37 and QTRG3 are pairwise aligned using ProgressiveCactus[25]. Then, using the pairwise global alignment of the two chromosomes, a list GRCh37 coordinates are converted to QTRG3 coordinates using HalTools[26] Given that the SNPs lifted over are present in dbSNP, the liftover success could be assessed using the dbSNP flanking sequenced for each SNP, which can be aligned to QTRG3 in order to identify a precise location using MUMer[27].

**Variant Functional Interpretation**

Functional interpretation was conducted for the 26 million SNPs and indels identified in Qatar using a combination of automated and manual algorithms. The automated curation consisted of identifying potentially deleterious coding variants, annotation and subdivision of the potentially deleterious variants based on presence or absence of the variants and/or affected genes in databases of variants and genes previously linked to phenotypes, followed by further subdivision based on variant allele frequency in Qatar.

Variants were assigned to genes in the ENSEMBL database build 75[28] using SnpEff 4.1[29]. In cases where multiple alternative transcripts overlap with distinct potential functions, the most severely deleterious function was reported, based on a prioritization hierarchy described in the annotation tool manual[29]. SnpEff classifies variant (SNP and indel) impact into three categories "high", "moderate", "low", or "modifier", where loss of function (LoF) variants are classified as high impact, nonsynonymous variants are classified as moderate impact, the remaining coding variants are classified as low impact, and most non-coding variants are classified as modifier impact. The high impact and moderate impact variants are considered potentially deleterious, and further classified with respect to databases of variant function and allele frequency. A known or potential link to a phenotype was determined using a database that combines Online Mendelian Inheritance in Man (OMIM)[30], the Human Gene Mutation Database (HGMD)[31], GWAS Catalog (GWAS)[32], and Pharmacogenomics Knowledgebase (PharmGKB)[32,33], ClinVar[34], and Human Phenotype Ontology (HPO)[35]. While the HGMD, GWAS, PharmGKB, and ClinVar databases provide annotation for variants, the OMIM database also provides information for genes and phenotypes, and HPO provides additional detail on the OMIM phenotypes (such as recessive/dominant classification). Qatari variants linked to a phenotype in the combined database of genes and variants linked to disease were identified based on the gene symbol containing the variant or the dbSNP rsID of the variant. The workflow for automated annotation starts with a VCF of all variants observed in the n=88 unrelated Qatari genomes or n=917 unrelated Qatari exomes. Each variant is assigned a function by SnpEff, with additional details provided for coding SNPs, such as transcript and polypeptide position and amino acid changes in HGVS format[36]. The variants were also assigned dbSNP 146 rsIDs, which were used to lookup the variant in the databases.

Potentially deleterious variants (high or moderate impact) are then queried against a database of variants linked to disease. The database includes the rsID, affected gene, phenotype linked to the variant, details of the phenotype, MIM ID of the gene and phenotype if available, and reference to the source database. This query produces the first group of variants (Category 1), where the variant (and the gene) are previously linked to a phenotype in one or more of the databases. The variants not present in the database are then queried against a list of all genes in the database, in order to produce the second group of variants (Category 2), variants not previously linked to a phenotype in genes previously linked to a phenotype. The remaining variants were placed a third group (Category 3), consisting of variants not linked to a phenotype in genes not previously linked to a phenotype.

The functional analysis was further stratified based on allele frequency. Each of the four categories of variants (all, Category 1, Category 2, Category 3) were stratified into 3 major groups (all, major reference allele, major alternate allele), and 5 minor groups (major reference allele / rare alternate allele, major reference allele / common alternate allele, major alternate alleles / common reference allele, major alternate allele / rare reference allele, major alternate allele / unobserved reference allele), where the threshold for "rare" alleles is 5% minor allele frequency and the threshold for "major" alleles is 50% major/minor allele frequency.

**Curated List of Known Pathogenic Variants for Mendelian Disorders in Qatar**

The set of Category 1 variants were filtered using a set of strict automated and manual curation steps, in order to produce a list of potential disease causing variants of interest for further study and potentially of value for an expanded genetic screening panel in Qatar[4]. The total set of potentially deleterious variants were filtered based on prevalence of the variant as well as the quality of literature linking the variant to a disease phenotype. The American College of Medical Genetics (ACMG) recommends excluding variants at prevalence above 5% minor allele frequency, in order to enrich for pathogenic variants[37]. Such a filter is consistent with a basic principle of evolutionary biology states that the prevalence of an allele in the population is expected to be inversely proportional to its pathogenic severity[38], with the most severely pathogenic variants expected to be rare. In addition, in order to minimize false positives, variants that were observed in at least two Qataris were keep for further consideration.

The Category 1 variants include all variants linked to a phenotype, including anthropomorphic traits (eye, hair color), GWAS hits, pharmacogenomics SNPs, and variants linked to Mendelian disorders, across the allele frequency spectrum. For the purpose of improving precision medicine in Qatar with respect to Mendelian disease variant screening, the focus of manual curation was on variants linked to Mendelian disorders. In order to identify a high-confidence set of known pathogenic variants, the database records and literature of the initial set of Category 1 variants linked to phenotypes were manually inspected. The initial criteria for extracting variants for manual curation were based on the phenotype (Mendelian or not). Variants were excluded from the manual curation list based on phenotype not listed as "recessive" or "dominant" Mendelian disorder in OMIM[30] and HPO[39]. For the remaining variants, a manual curation process was implemented to identify a subset of high-confidence pathogenic variants for precision medicine disease screening in Qatar. First, the entry for each variant on the OMIM webpage was evaluated to determine the strength of evidence supporting this variant's pathogenicity in the reported disease. For this step, the entry was scrutinized by a geneticist blinded to variant-specific metrics (observed homozygotes/heterozygotes), in each instance reporting inheritance mode of the disease documented for this variant (as RECESSIVE, DOMINANT, SEMIDOMINANT, DIGENIC or MODIFIER) along with relevant severity and onset information (as MILD, LATE-ONSET, INCOMPLETE/REDUCED PENTRANCE, or PHARMACOGENETIC). Where any of this information was missing from the OMIM page, the primary literature was investigated in detail by reviewing the initial publication in which the subject with disease was initially described, as well as any subsequent publication that supported or overturned the pathogenicity. The final criteria for exclusion was variants associated with but not causal for a disease (conferring odds of risk of disease), pharmacogenetic variants (requiring a challenge by a drug for a phenotype), or variants that were reclassified in more recent studies as polymorphisms or variants of unknown significance.

**Supplemental Table I. Batch-specific variant filters for call set integration of 108 genomes and 1268 exomes** [1]

| Nickname | Genome | | Exome 38 Mb | | Exome 51 Mb | |
|---|---|---|---|---|---|---|
| Sequencing breadth | Whole genome | | Whole exome | | Whole exome | |
| Sequencing year | 2012 | | 2010 | | 2014 | |
| Enrichment platform | N/A | | SureSelect 38 Mb | | SureSelect 51 Mb | |
| Sequencing platform | Illumina HiSeq 2500 | | Illumina HiSeq 2000 | | Illumina HiSeq 2500 | |
| Sample size | 108 | | 67 | | 1201 | |
| **Filters** | | | | | | |
| Minimum depth | 12 | | 12 | | 10 | |
| Minimum allele count | 1 | | 1 | | 2 | |
| Mean depth | 38 | | 70 | | 70 | |
| Novel variants (% not in dbSNP 146) | 0.73% | | 0.72% | | 0.72% | |
| **Autosomal variant summary** | **All** | **Novel** | **All** | **Novel** | **All** | **Novel** |
| Variant alleles | 19,355 | 104 | 18,537 | 100 | 19,229 | 103 |
| Variant sites | 14,107 | 103 | 13,682 | 98 | 14,064 | 102 |
| Reference homozygotes | 287,672 | 62,601 | 268,401 | 58,600 | 161,239 | 19,421 |
| Heterozygotes | 8,860 | 102 | 8,826 | 97 | 8,898 | 101 |
| Variant homozygotes | 5,247 | 1 | 4,856 | 1 | 5,166 | 1 |
| Homozygotes | 292,919 | 62,603 | 273,257 | 58,602 | 166,405 | 19,422 |

[1.] In order to produce an integrated variant call set based on sequence data from three different batches (1 genome, 2 exome), batch-specific variant filters were applied, such that the novel SNP rate was consistent across platforms when inspecting genomic intervals covered by all platforms. Genotypes generated using GATK best practices were produced for autosomal coding exons covered by Agilent SureSelect 38 Mb targets, the most restricted platform used (with respect to breadth of genome sampled), and further limited to known coding exons present in the CCDS database (which is an intersection of RefSeq, ENSEMBL, and UCSC defined genes). From the initial call set, sites with a PHRED-scaled genotype quality above 30 were further filtered using a range of minimum depth and minimum allele count, and compared with respect to novel SNP rate (defined as the percentage of variant sites not previously observed in dbSNP build 146). The combinations were sorted by increasing novel SNP rate, and batch-specific parameters were selected such that (a) both exome platform novel SNP rates are lower than the genome novel SNP rate, the genome novel SNP rate is as close as possible to the exome novel SNP rate, and the lowest SNP rate is as high as possible. Based on analysis of a quadruple-sequenced Qatari, the batch-specific novel variant rate for these parameters is below 5%. Shown are the sequencing parameters, calibration parameters, and autosomal variant summary for each platform. Sequencing parameters include a name for each platform (including the sequencing center), the sequencing breadth (exome or genome), the year the sequencing was completed, the target enrichment platform, the sequencing platform. The Calibration parameters include the minimum depth, minimum allele count, mean variant depth, and percentage of novel variants (not in dbSNP 146) for the selected optimal batch-specific filters. The autosomal variant summary presents for all and for novel SNPs the average result per individual for the selected calibration parameters, including number of variant alleles, variant sites, reference homozygous genotypes, heterozygous genotypes, variant homozygous genotypes, and total homozygotes.

**Supplemental Table II. Populations included in ADMIXTURE and PCA analysis[1]**

| Study | Region | Population | n |
|---|---|---|---|
| 1000Gp1 | AFR | ASW | 61 |
| 1000Gp1 | AFR | LWK | 95 |
| 1000Gp1 | AFR | YRI | 57 |
| 1000Gp1 | AMR | CLM | 60 |
| 1000Gp1 | AMR | MXL | 66 |
| 1000Gp1 | AMR | PUR | 55 |
| 1000Gp1 | EAS | CHB | 97 |
| 1000Gp1 | EAS | CHS | 100 |
| 1000Gp1 | EAS | JPT | 89 |
| 1000Gp1 | EUR | CEU | 85 |
| 1000Gp1 | EUR | FIN | 86 |
| 1000Gp1 | EUR | GBR | 76 |
| 1000Gp1 | EUR | IBS | 3 |
| 1000Gp1 | EUR | TSI | 98 |
| 1000Gp3 | AMR | ACB | 96 |
| 1000Gp3 | AMR | ASW | 7 |
| 1000Gp3 | AMR | CLM | 36 |
| 1000Gp3 | AMR | MXL | 5 |
| 1000Gp3 | AMR | PEL | 85 |
| 1000Gp3 | AMR | PUR | 49 |
| 1000Gp3 | EASN | CDX | 93 |
| 1000Gp3 | EASN | CHB | 8 |
| 1000Gp3 | EASN | CHS | 13 |
| 1000Gp3 | EASN | JPT | 15 |
| 1000Gp3 | EASN | KHV | 91 |
| 1000Gp3 | EUR | CEU | 15 |
| 1000Gp3 | EUR | FIN | 10 |
| 1000Gp3 | EUR | GBR | 9 |
| 1000Gp3 | EUR | IBS | 44 |
| 1000Gp3 | EUR | TSI | 11 |
| 1000Gp3 | SASN | BEB | 79 |
| 1000Gp3 | SASN | GIH | 86 |
| 1000Gp3 | SASN | ITU | 102 |
| 1000Gp3 | SASN | PJL | 88 |
| 1000Gp3 | SASN | STU | 102 |
| 1000Gp3 | WAFR | ESN | 91 |
| 1000Gp3 | WAFR | GWD | 107 |
| 1000Gp3 | WAFR | LWK | 12 |
| 1000Gp3 | WAFR | MSL | 78 |
| 1000Gp3 | WAFR | YRI | 13 |
| Non-Qatari in Qatar Exome | Africa | Algerian | 1 |
| Non-Qatari in Qatar Exome | Africa | American/Egyptian | 5 |
| Non-Qatari in Qatar Exome | Africa | Egyptian | 41 |
| Non-Qatari in Qatar Exome | Africa | Eritrean | 4 |

**Supplemental Table II. Populations included in ADMIXTURE and PCA analysis[1] (cont., page 2)**

| Study | Region | Population | n |
|---|---|---|---|
| Non-Qatari in Qatar Exome | Africa | Morracan | 1 |
| Non-Qatari in Qatar Exome | Africa | Nigerian | 2 |
| Non-Qatari in Qatar Exome | America | American | 5 |
| Non-Qatari in Qatar Exome | America | Argentenian | 1 |
| Non-Qatari in Qatar Exome | East Asia | Filipino | 3 |
| Non-Qatari in Qatar Exome | East Asia | Nepalese | 1 |
| Non-Qatari in Qatar Exome | Middle East | Arab | 1 |
| Non-Qatari in Qatar Exome | Middle East | Iranian | 10 |
| Non-Qatari in Qatar Exome | Middle East | Iraqi | 1 |
| Non-Qatari in Qatar Exome | Middle East | Jordanian | 21 |
| Non-Qatari in Qatar Exome | Middle East | Jordanian/Arab | 3 |
| Non-Qatari in Qatar Exome | Middle East | Jordanian/Palestinian | 1 |
| Non-Qatari in Qatar Exome | Middle East | Lebanese | 4 |
| Non-Qatari in Qatar Exome | Middle East | Omani | 8 |
| Non-Qatari in Qatar Exome | Middle East | Palestinian | 20 |
| Non-Qatari in Qatar Exome | Middle East | Saudi | 3 |
| Non-Qatari in Qatar Exome | Middle East | Sri-Lankan | 1 |
| Non-Qatari in Qatar Exome | Middle East | Sudanese | 16 |
| Non-Qatari in Qatar Exome | Middle East | Syrian | 9 |
| Non-Qatari in Qatar Exome | Middle East | Tunisian | 5 |
| Non-Qatari in Qatar Exome | Middle East | Yemeni | 5 |
| Non-Qatari in Qatar Exome | South Asia | Bangladeshi | 2 |
| Non-Qatari in Qatar Exome | South Asia | Indian | 10 |
| Non-Qatari in Qatar Exome | South Asia | Indian/Asian | 5 |
| Non-Qatari in Qatar Exome | South Asia | Pakistani | 26 |
| Human Origins | Africa | AA Denver | 12 |
| Human Origins | Africa | Algerian | 7 |
| Human Origins | Africa | Bantu SA Herero | 2 |
| Human Origins | Africa | Bantu SA Ovambo | 1 |
| Human Origins | Africa | Bantu SA Pedi | 1 |
| Human Origins | Africa | Bantu SA S Sotho | 1 |
| Human Origins | Africa | Bantu SA Tswana | 2 |
| Human Origins | Africa | Bantu SA Zulu | 1 |
| Human Origins | Africa | BantuKenya | 6 |
| Human Origins | Africa | BiakaPygmy | 20 |
| Human Origins | Africa | Datog | 3 |
| Human Origins | Africa | Esan Nigeria ESN | 8 |
| Human Origins | Africa | Gambian GWD | 6 |
| Human Origins | Africa | Hadza Henn | 5 |
| Human Origins | Africa | Ju hoan North | 5 |
| Human Origins | Africa | Khomani | 11 |
| Human Origins | Africa | Kikuyu | 4 |
| Human Origins | Africa | Luhya Kenya LWK | 8 |
| Human Origins | Africa | Luo | 8 |
| Human Origins | Africa | Mandenka | 17 |

**Supplemental Table II. Populations included in ADMIXTURE and PCA analysis[1] (cont., page 3)**

| Study | Region | Population | n |
|---|---|---|---|
| Human Origins | Africa | Masai Ayodo | 2 |
| Human Origins | Africa | Masai Kinyawa MKK | 10 |
| Human Origins | Africa | MbutiPygmy | 10 |
| Human Origins | Africa | Mende Sierra Leone MSL | 8 |
| Human Origins | Africa | Mozabite | 21 |
| Human Origins | Africa | Saharawi | 6 |
| Human Origins | Africa | Somali | 13 |
| Human Origins | Africa | Tunisian | 8 |
| Human Origins | Africa | Yoruba | 70 |
| Human Origins | America | Bolivian Cochabamba | 1 |
| Human Origins | America | Bolivian LaPaz | 3 |
| Human Origins | America | Bolivian Pando | 3 |
| Human Origins | America | Karitiana | 12 |
| Human Origins | America | Mayan | 18 |
| Human Origins | America | Mixe | 10 |
| Human Origins | America | Mixtec | 10 |
| Human Origins | America | Piapoco | 4 |
| Human Origins | America | Pima | 14 |
| Human Origins | America | Quechua Coriell | 5 |
| Human Origins | America | Surui | 8 |
| Human Origins | America | Zapotec | 10 |
| Human Origins | Central Asia Siberia | Aleut | 7 |
| Human Origins | Central Asia Siberia | Altaian | 7 |
| Human Origins | Central Asia Siberia | Chukchi | 20 |
| Human Origins | Central Asia Siberia | Chukchi Reindeer | 1 |
| Human Origins | Central Asia Siberia | Chukchi Sir | 2 |
| Human Origins | Central Asia Siberia | Dolgan | 3 |
| Human Origins | Central Asia Siberia | Eskimo Chaplin | 4 |
| Human Origins | Central Asia Siberia | Eskimo Naukan | 13 |
| Human Origins | Central Asia Siberia | Eskimo Sireniki | 5 |
| Human Origins | Central Asia Siberia | Even | 10 |
| Human Origins | Central Asia Siberia | Itelmen | 6 |
| Human Origins | Central Asia Siberia | Kalmyk | 10 |
| Human Origins | Central Asia Siberia | Koryak | 9 |
| Human Origins | Central Asia Siberia | Kyrgyz | 9 |
| Human Origins | Central Asia Siberia | Mansi | 8 |
| Human Origins | Central Asia Siberia | Mongola | 6 |
| Human Origins | Central Asia Siberia | Nganasan | 11 |
| Human Origins | Central Asia Siberia | Selkup | 10 |
| Human Origins | Central Asia Siberia | Tajik Pomiri | 8 |
| Human Origins | Central Asia Siberia | Tlingit | 4 |
| Human Origins | Central Asia Siberia | Tubalar | 22 |
| Human Origins | Central Asia Siberia | Turkmen | 7 |
| Human Origins | Central Asia Siberia | Tuvinian | 10 |
| Human Origins | Central Asia Siberia | Ulchi | 25 |

**Supplemental Table II. Populations included in ADMIXTURE and PCA analysis[1] (cont., page 4)**

| Study | Region | Population | n |
|---|---|---|---|
| Human Origins | Central Asia Siberia | Uzbek | 10 |
| Human Origins | Central Asia Siberia | Yakut | 20 |
| Human Origins | Central Asia Siberia | Yukagir Forest | 5 |
| Human Origins | Central Asia Siberia | Yukagir Tundra | 14 |
| Human Origins | East Asia | Ami Coriell | 10 |
| Human Origins | East Asia | Atayal Coriell | 9 |
| Human Origins | East Asia | Cambodian | 8 |
| Human Origins | East Asia | Dai | 10 |
| Human Origins | East Asia | Daur | 9 |
| Human Origins | East Asia | Han | 33 |
| Human Origins | East Asia | Han NChina | 10 |
| Human Origins | East Asia | Hezhen | 8 |
| Human Origins | East Asia | Japanese | 29 |
| Human Origins | East Asia | Kinh Vietnam KHV | 8 |
| Human Origins | East Asia | Korean | 6 |
| Human Origins | East Asia | Lahu | 8 |
| Human Origins | East Asia | Miao | 10 |
| Human Origins | East Asia | Naxi | 9 |
| Human Origins | East Asia | Oroqen | 9 |
| Human Origins | East Asia | She | 10 |
| Human Origins | East Asia | Thai | 10 |
| Human Origins | East Asia | Tu | 10 |
| Human Origins | East Asia | Tujia | 10 |
| Human Origins | East Asia | Uygur | 10 |
| Human Origins | East Asia | Xibo | 7 |
| Human Origins | East Asia | Yi | 10 |
| Human Origins | Middle East | BedouinA | 25 |
| Human Origins | Middle East | BedouinB | 19 |
| Human Origins | Middle East | Druze | 39 |
| Human Origins | Middle East | Egyptian Comas | 11 |
| Human Origins | Middle East | Egyptian Metspalu | 7 |
| Human Origins | Middle East | Iranian | 8 |
| Human Origins | Middle East | Jordanian | 9 |
| Human Origins | Middle East | Lebanese | 8 |
| Human Origins | Middle East | Palestinian | 38 |
| Human Origins | Middle East | Saudi | 8 |
| Human Origins | Middle East | Syrian | 8 |
| Human Origins | Middle East | Turkish | 4 |
| Human Origins | Middle East | Turkish Adana | 10 |
| Human Origins | Middle East | Turkish Aydin | 7 |
| Human Origins | Middle East | Turkish Balikesir | 6 |
| Human Origins | Middle East | Turkish Istanbul | 10 |
| Human Origins | Middle East | Turkish Kayseri | 10 |
| Human Origins | Middle East | Turkish Trabzon | 9 |
| Human Origins | Middle East | Yemen | 6 |

**Supplemental Table II. Populations included in ADMIXTURE and PCA analysis[1] (cont., page 5)**

| Study | Region | Population | n |
|---|---|---|---|
| Human Origins | Oceania | Australian ECCAC | 3 |
| Human Origins | Oceania | Bougainville | 10 |
| Human Origins | Oceania | Papuan | 14 |
| Human Origins | South Asia | Balochi | 20 |
| Human Origins | South Asia | Bengali Bangladesh BEB | 7 |
| Human Origins | South Asia | Brahui | 21 |
| Human Origins | South Asia | Burusho | 23 |
| Human Origins | South Asia | GujaratiA GIH | 5 |
| Human Origins | South Asia | GujaratiB GIH | 5 |
| Human Origins | South Asia | GujaratiC GIH | 5 |
| Human Origins | South Asia | GujaratiD GIH | 5 |
| Human Origins | South Asia | Hazara | 14 |
| Human Origins | South Asia | Kalash | 18 |
| Human Origins | South Asia | Kusunda | 10 |
| Human Origins | South Asia | Makrani | 20 |
| Human Origins | South Asia | Pathan | 19 |
| Human Origins | South Asia | Punjabi Lahore PJL | 8 |
| Human Origins | South Asia | Sindhi | 18 |
| Human Origins | West Eurasia | Abkhasian | 9 |
| Human Origins | West Eurasia | Adygei | 17 |
| Human Origins | West Eurasia | Albanian | 6 |
| Human Origins | West Eurasia | Armenian | 10 |
| Human Origins | West Eurasia | Balkar | 10 |
| Human Origins | West Eurasia | Basque French | 20 |
| Human Origins | West Eurasia | Basque Spanish | 9 |
| Human Origins | West Eurasia | Belarusian | 10 |
| Human Origins | West Eurasia | Bulgarian | 10 |
| Human Origins | West Eurasia | Chechen | 9 |
| Human Origins | West Eurasia | Chuvash | 10 |
| Human Origins | West Eurasia | Croatian | 10 |
| Human Origins | West Eurasia | Cypriot | 8 |
| Human Origins | West Eurasia | Czech | 10 |
| Human Origins | West Eurasia | English Cornwall GBR | 5 |
| Human Origins | West Eurasia | English Kent GBR | 5 |
| Human Origins | West Eurasia | Estonian | 10 |
| Human Origins | West Eurasia | Finnish FIN | 7 |
| Human Origins | West Eurasia | French | 25 |
| Human Origins | West Eurasia | French South | 7 |
| Human Origins | West Eurasia | Georgian Megrels | 10 |
| Human Origins | West Eurasia | Greek Comas | 14 |
| Human Origins | West Eurasia | Greek Coriell | 6 |
| Human Origins | West Eurasia | Hungarian Coriell | 10 |
| Human Origins | West Eurasia | Hungarian Metspalu | 10 |
| Human Origins | West Eurasia | Icelandic | 12 |
| Human Origins | West Eurasia | Italian Bergamo | 12 |

**Supplemental Table II. Populations included in ADMIXTURE and PCA analysis[1] (cont., page 6)**

| Study | Region | Population | n |
|---|---|---|---|
| Human Origins | West Eurasia | Italian EastSicilian | 5 |
| Human Origins | West Eurasia | Italian South | 1 |
| Human Origins | West Eurasia | Italian Tuscan | 8 |
| Human Origins | West Eurasia | Italian WestSicilian | 6 |
| Human Origins | West Eurasia | Kumyk | 8 |
| Human Origins | West Eurasia | Lezgin | 9 |
| Human Origins | West Eurasia | Lithuanian | 10 |
| Human Origins | West Eurasia | Maltese | 8 |
| Human Origins | West Eurasia | Mordovian | 10 |
| Human Origins | West Eurasia | Nogai | 9 |
| Human Origins | West Eurasia | North Ossetian | 10 |
| Human Origins | West Eurasia | Norwegian | 11 |
| Human Origins | West Eurasia | Orcadian | 13 |
| Human Origins | West Eurasia | Russian | 22 |
| Human Origins | West Eurasia | Saami WGA | 1 |
| Human Origins | West Eurasia | Sardinian | 27 |
| Human Origins | West Eurasia | Scottish Argyll Bute GBR | 4 |
| Human Origins | West Eurasia | Spanish Andalucia IBS | 4 |
| Human Origins | West Eurasia | Spanish Aragon IBS | 6 |
| Human Origins | West Eurasia | Spanish Baleares IBS | 4 |
| Human Origins | West Eurasia | Spanish Canarias IBS | 2 |
| Human Origins | West Eurasia | Spanish Cantabria IBS | 5 |
| Human Origins | West Eurasia | Spanish Castilla la Mancha IBS | 5 |
| Human Origins | West Eurasia | Spanish Castilla y Leon IBS | 5 |
| Human Origins | West Eurasia | Spanish Cataluna IBS | 5 |
| Human Origins | West Eurasia | Spanish Extremadura IBS | 5 |
| Human Origins | West Eurasia | Spanish Galicia IBS | 5 |
| Human Origins | West Eurasia | Spanish Murcia IBS | 4 |
| Human Origins | West Eurasia | Spanish Pais Vasco IBS | 5 |
| Human Origins | West Eurasia | Spanish Valencia IBS | 5 |
| Human Origins | West Eurasia | Ukrainian East | 6 |
| Human Origins | West Eurasia | Ukrainian West | 3 |
| Qatar Exome | Qatar | Qatari | 108 |
| Qatar Genome | Qatar | Qatari | 1268 |

[1] Shown is the sample size in the dataset used for population structure analysis, including samples from 1000 Genomes Phase 1[40] (1000Gp1), 1000 Genomes Phase 3[41] (1000Gp3), Human Origins (HumOr), Qatar Genome (QG), Qatar Exome (QE), and Non-Qataris in Qatar Exome (NQE). A number of samples overlap between HumOr, 1000Gp1, and 1000Gp3, in these cases firs the HumOr samples are added, followed by non-duplicate samples in 1000Gp3, and finally non-duplicate samples in 1000Gp3. In order to distinguish between 1000G and HumOr samples for the same populations, the 1000 Genomes populations are described by a 3-letter code (such as YRI), while HumOr samples use the full name (such as Yoruba) in Lazardis et al 2014 supplement[42]. Shown is the study, region, population, and sample size. For Qataris, the sample size is based the sequencing technology used (genome or exome). Regions are based on designations in Lazardis et al 2014 (Africa, America, East Asia, Middle East, Qatar, South Asia, Central Asia / Siberia, Oceania, West Eurasia) and the 1000 Genomes website (AFR, AMR, EAS, EUR, EASN, SASN, WAFR), where 3-letter codes are used for Phase 1 and 4-letter codes are used for Phase 3.

## Supplemental Table III. Population structure analysis[1]

| Study | n | K (ADMIXTURE K=12) | | | | | | | | | | | |
|---|---|---|---|---|---|---|---|---|---|---|---|---|---|
| | | 1 | 2 | 3 | 4 | 5 | 6 | 7 | 8 | 9 | 10 | 11 | 12 |
| Inferred origin | | European | East Asian | Siberian | South Asian | Bedouin | African Pygmy | Oceanian | Arab | Persian | Sub-Saharan African | American | Central Asian |
| Qatar Genome | 108 | 0 | 0 | 0 | 4 | 48 | 0 | 0 | 20 | 19 | 17 | 0 | 0 |
| Qatar Exome | 1053 | 5 | 0 | 0 | 78 | 518 | 1 | 0 | 216 | 175 | 60 | 0 | 0 |
| in Qatar Exome | 215 | 11 | 2 | 0 | 49 | 35 | 2 | 0 | 9 | 94 | 13 | 0 | 0 |
| Human Origins | 1862 | 418 | 195 | 103 | 176 | 65 | 51 | 27 | 0 | 353 | 184 | 100 | 190 |
| 1000Gp1 | 1028 | 391 | 238 | 0 | 0 | 0 | 0 | 0 | 0 | 53 | 213 | 85 | 48 |
| 1000Gp3 | 1345 | 134 | 212 | 0 | 457 | 0 | 0 | 0 | 0 | 23 | 408 | 103 | 8 |
| Public data | 4235 | 943 | 645 | 103 | 633 | 65 | 51 | 27 | 0 | 429 | 805 | 288 | 246 |
| Qatar | 1161 | 5 | 0 | 0 | 82 | 566 | 1 | 0 | 236 | 194 | 77 | 0 | 0 |
| Total | 2373 | 525 | 450 | 0 | 457 | 0 | 0 | 0 | 0 | 76 | 621 | 188 | 56 |

[1] Summary of population structure analysis using ADMIXTURE. Genotypes were obtained for 2265 SNPs in a combined set of 5611 individuals, including 1161 Qataris (108 genomes and 1053 exomes), 215 s living in Qatar, 1862 from Human Origins dataset[42], 1028 from 1000 Genomes Phase 1 not in Human Origins[40], and 1345 from 1000 Genomes Phase 3 not in Human Origins nor 1000 Genomes Phase 1. The set of SNPs used for the analysis were segregating in all three datasets, and were pruned using PLINK2[13] to exclude SNPs with Hardy-Weinberg equilibrium p-value $< 1 \times 10^{-6}$ and SNPs with linkage disequilibrium $> 0.25$ in 1000 SNP intervals (step size 25 SNPs). ADMIXTURE analysis was conducted K=12, based on lowest cross-validation error in a prior study[16,17]. The resulting ancestry proportions in 12 ancestral populations for each individual was used to assign each individual to a cluster for color-coding principal components analysis, based on the dominant ancestry for each individual. Shown is the total sample and number of individuals assigned to each of 12 clusters (K) for Qatari Exomes, Qatari Genomes, s in Qatar, Human Origins, 1000 Genomes Phase 1, 1000 Genomes Phase 3, and totals for public data (Human Origins, 1000 Genomes Phase 1 and 1000 Genomes Phase 3) and Qatari data (Qatar Genome and Qatar Exome). Cluster names "inferred origin" is based on the major populations present in the cluster.

**Supplemental Table IV. Relatives Identified in 1376 Qatari and Non-Qatari Exomes[1]**

| Sample | | Inferred relation | Kept | Pairs (n) |
| --- | --- | --- | --- | --- |
| 1st | 2nd | | | |
| Qatari | Qatari | 3 | both | 294 |
| Qatari | Qatari | 1 | one | 110 |
| Qatari | Qatari | 3 | one | 106 |
| Non-Qatari | Non-Qatari | 1 | neither | 87 |
| Qatari | Qatari | 2 | one | 55 |
| Qatari | Qatari | 1 | neither | 38 |
| Non-Qatari | Non-Qatari | 3 | both | 10 |
| Qatari | Qatari | 3 | neither | 9 |
| Qatari | Qatari | 2 | neither | 5 |
| Non-Qatari | Non-Qatari | 2 | one | 4 |
| Non-Qatari | Non-Qatari | 3 | one | 4 |
| Qatari | Qatari | 2 | both | 4 |
| Qatari | Non-Qatari | 1 | neither | 3 |
| Non-Qatari | Non-Qatari | 2 | neither | 2 |
| Qatari | Non-Qatari | 3 | one | 2 |
| Qatari | Non-Qatari | 1 | one | 1 |
| Qatari | Non-Qatari | 2 | one | 1 |
| Qatari | Non-Qatari | 3 | both | 1 |
| Non-Qatari | Non-Qatari | 1 | both | 0 |
| Non-Qatari | Non-Qatari | 2 | both | 0 |
| Non-Qatari | Non-Qatari | 3 | neither | 0 |
| Qatari | Non-Qatari | 1 | both | 0 |
| Qatari | Non-Qatari | 2 | both | 0 |
| Qatari | Non-Qatari | 2 | neither | 0 |
| Qatari | Non-Qatari | 3 | neither | 0 |
| Qatari | Qatari | 1 | both | 0 |

[1] SNPs were pruned from the initial 1376 individuals' exome datasets of 767,274 SNPs based on surrounding linkage disequilibrium (LD) structure using PLINK 1.90's --indep-pairwise option with window sizes of 1000 SNPs, overlaps of 25 SNPs, and SNPs with LD > 0.25 pruned out. This left 381,028 SNPs for analysis. Manichaikul et al.'s[19] program for estimating kinship coefficients between pairs of individuals, KING-robust v1.4, was then applied to this filtered dataset using the flag --kinship. None of the known pedigree information was included in the KING analysis. Degree of relatedness was inferred via cutoffs provided in the Manichaikul et al.[19] paper, except for our second degree lower bound value (0.1101) which was chosen based on an analysis of SNP data from 1000 Genomes and HapMap on individuals with known relationships. Columns 1 and 2 of the table denote whether individual 1 and two are currently living in Qatar or are Non-Qataris, respectively. Column 3 is the relationship inferred by the KING-robust analysis. Column 4 denotes whether both, one, or neither individual was included in the subsequent analyses. Column 5 provides the number of pairs for which the prior columns' information applies to.

**Supplemental Table V. Aggregated variants observed in 1005 unrelated Qatari (88 genomes, 917 exomes) and incorporated into the Qatar Reference Genome[1]**

| | SNPs | | | | | | | | | | | | | | | | | | | | Indels | | | | |
|---|---|---|---|---|---|---|---|---|---|---|---|---|---|---|---|---|---|---|---|---|---|---|---|---|---|
| | Integrated (n=1005) | | | | | Genome (n=88) | | | | | Exome 50 Mb (n=853) | | | | | Exome 38 Mb (n=64) | | | | | Genome (n=88) | | | | |
| | | Novel | | Major alt allele | | | Novel | | Major alt allele | | | Novel | | Major alt allele | | | Novel | | Major alt allele | | | Novel | | Major alt allele | |
| Chr | SNPs | N | % | N | % | SNPs | N | % | N | % | SNPs | N | % | N | % | SNPs | N | % | N | % | Indels | N | % | N | % |
| All | 20,937,965 | 2,975,232 | 14.21% | 1,931,122 | 9.22% | 20,838,516 | 2,956,681 | 14.19% | 1,931,067 | 9.27% | 167,723 | 18,175 | 10.84% | 7,567 | 4.51% | 84,133 | 5,973 | 7.10% | 7,065 | 8.40% | 5,452,613 | 3,182,624 | 58.37% | 1,882,405 | 34.52% |
| 1 | 1,592,163 | 222,736 | 13.99% | 149,248 | 9.37% | 1,581,236 | 220,786 | 13.96% | 149,224 | 9.44% | 18,537 | 1,886 | 10.17% | 860 | 4.64% | 9,392 | 615 | 6.55% | 781 | 8.32% | 423,410 | 248,009 | 58.57% | 143,511 | 33.89% |
| 2 | 1,709,761 | 240,085 | 14.04% | 154,410 | 9.03% | 1,703,529 | 238,883 | 14.02% | 154,421 | 9.06% | 10,370 | 1,155 | 11.14% | 489 | 4.72% | 5,296 | 383 | 7.23% | 458 | 8.65% | 439,507 | 255,079 | 58.04% | 148,206 | 33.72% |
| 3 | 1,409,619 | 187,933 | 13.33% | 129,413 | 9.18% | 1,404,001 | 186,838 | 13.31% | 129,414 | 9.22% | 9,227 | 1,035 | 11.22% | 400 | 4.34% | 4,695 | 358 | 7.63% | 385 | 8.20% | 357,315 | 206,465 | 57.78% | 120,288 | 33.66% |
| 4 | 1,445,319 | 195,681 | 13.54% | 144,549 | 10.00% | 1,441,704 | 195,004 | 13.53% | 144,551 | 10.03% | 6,261 | 621 | 9.92% | 305 | 4.87% | 3,286 | 246 | 7.49% | 291 | 8.86% | 370,841 | 217,409 | 58.63% | 125,936 | 33.96% |
| 5 | 1,282,399 | 178,124 | 13.89% | 114,208 | 8.91% | 1,278,212 | 177,320 | 13.87% | 114,208 | 8.93% | 7,061 | 784 | 11.10% | 388 | 5.49% | 3,546 | 256 | 7.22% | 366 | 10.32% | 320,147 | 182,821 | 57.11% | 109,029 | 34.06% |
| 6 | 1,254,721 | 169,944 | 13.54% | 119,019 | 9.49% | 1,249,428 | 168,993 | 13.53% | 119,020 | 9.53% | 9,406 | 1,143 | 12.15% | 447 | 4.75% | 5,019 | 535 | 10.66% | 409 | 8.15% | 332,004 | 195,950 | 59.02% | 114,112 | 34.37% |
| 7 | 1,181,277 | 163,189 | 13.81% | 105,329 | 8.92% | 1,176,840 | 162,335 | 13.79% | 105,324 | 8.95% | 7,368 | 846 | 11.48% | 262 | 3.56% | 3,689 | 271 | 7.35% | 261 | 7.08% | 310,413 | 181,560 | 58.49% | 103,381 | 33.30% |
| 8 | 1,121,946 | 154,901 | 13.81% | 100,654 | 8.97% | 1,118,463 | 154,271 | 13.79% | 100,653 | 9.00% | 5,741 | 620 | 10.80% | 225 | 3.92% | 2,778 | 185 | 6.66% | 215 | 7.74% | 263,130 | 153,087 | 58.18% | 87,994 | 33.44% |
| 9 | 893,371 | 124,547 | 13.94% | 81,215 | 9.09% | 889,068 | 123,746 | 13.92% | 81,222 | 9.14% | 7,154 | 778 | 10.88% | 290 | 4.05% | 3,493 | 237 | 6.78% | 270 | 7.73% | 220,843 | 126,445 | 57.26% | 74,180 | 33.59% |
| 10 | 1,011,674 | 132,809 | 13.13% | 94,540 | 9.34% | 1,007,822 | 132,100 | 13.11% | 94,529 | 9.38% | 6,692 | 663 | 9.91% | 315 | 4.71% | 3,453 | 229 | 6.63% | 302 | 8.75% | 262,781 | 152,899 | 58.18% | 88,800 | 33.79% |
| 11 | 992,815 | 133,836 | 13.48% | 101,238 | 10.20% | 986,571 | 132,732 | 13.45% | 101,237 | 10.26% | 10,547 | 1,100 | 10.43% | 568 | 5.39% | 5,359 | 328 | 6.12% | 537 | 10.02% | 246,040 | 142,728 | 58.01% | 82,771 | 33.64% |
| 12 | 950,274 | 127,170 | 13.38% | 89,829 | 9.45% | 945,483 | 126,273 | 13.36% | 89,819 | 9.50% | 8,066 | 871 | 10.80% | 378 | 4.69% | 4,078 | 280 | 6.87% | 360 | 8.83% | 258,803 | 149,851 | 57.90% | 87,556 | 33.83% |
| 13 | 721,704 | 94,831 | 13.14% | 75,313 | 10.44% | 720,085 | 94,533 | 13.13% | 75,309 | 10.46% | 2,783 | 291 | 10.46% | 148 | 5.32% | 1,457 | 110 | 7.55% | 141 | 9.68% | 192,835 | 115,031 | 59.65% | 66,223 | 34.34% |
| 14 | 657,214 | 90,851 | 13.82% | 60,225 | 9.16% | 654,196 | 90,281 | 13.80% | 60,221 | 9.21% | 5,139 | 585 | 11.38% | 243 | 4.73% | 2,603 | 163 | 6.26% | 236 | 9.07% | 169,405 | 97,754 | 57.70% | 58,262 | 34.39% |
| 15 | 605,125 | 86,523 | 14.30% | 55,833 | 9.23% | 602,017 | 85,933 | 14.27% | 55,831 | 9.27% | 5,182 | 571 | 11.02% | 233 | 4.50% | 2,624 | 180 | 6.86% | 225 | 8.57% | 157,232 | 91,850 | 58.42% | 54,406 | 34.60% |
| 16 | 667,442 | 97,950 | 14.68% | 57,149 | 8.56% | 662,638 | 97,058 | 14.65% | 57,150 | 8.62% | 7,894 | 882 | 11.17% | 275 | 3.48% | 3,630 | 270 | 7.44% | 245 | 6.75% | 161,821 | 94,228 | 58.23% | 52,265 | 32.30% |
| 17 | 573,728 | 83,775 | 14.60% | 49,208 | 8.58% | 567,883 | 82,664 | 14.56% | 49,198 | 8.66% | 9,903 | 1,074 | 10.85% | 422 | 4.26% | 4,821 | 330 | 6.85% | 388 | 8.05% | 165,430 | 96,122 | 58.10% | 55,824 | 33.74% |
| 18 | 567,901 | 74,034 | 13.04% | 55,536 | 9.78% | 566,489 | 73,776 | 13.02% | 55,537 | 9.80% | 2,472 | 248 | 10.03% | 121 | 4.89% | 1,290 | 85 | 6.59% | 125 | 9.69% | 146,951 | 85,698 | 58.32% | 50,757 | 34.54% |
| 19 | 478,025 | 66,775 | 13.97% | 39,669 | 8.30% | 470,516 | 65,330 | 13.88% | 39,674 | 8.43% | 12,734 | 1,441 | 11.32% | 499 | 3.92% | 5,900 | 366 | 6.20% | 449 | 7.61% | 141,734 | 81,136 | 57.25% | 47,778 | 33.71% |
| 20 | 456,002 | 58,746 | 12.88% | 36,920 | 8.10% | 453,008 | 58,231 | 12.85% | 36,917 | 8.15% | 5,099 | 475 | 9.32% | 221 | 4.33% | 2,482 | 151 | 6.08% | 192 | 7.74% | 120,911 | 72,303 | 59.80% | 39,609 | 32.76% |
| 21 | 297,875 | 42,861 | 14.39% | 30,364 | 10.19% | 296,542 | 42,658 | 14.39% | 30,363 | 10.24% | 2,316 | 193 | 8.33% | 124 | 5.35% | 1,141 | 58 | 5.08% | 111 | 9.73% | 87,587 | 54,374 | 62.08% | 26,897 | 30.71% |
| 22 | 289,339 | 39,574 | 13.68% | 24,499 | 8.47% | 286,767 | 39,142 | 13.65% | 24,494 | 8.54% | 4,404 | 432 | 9.81% | 187 | 4.25% | 2,082 | 118 | 5.67% | 168 | 8.07% | 76,727 | 46,370 | 60.44% | 25,186 | 32.83% |
| X | 770,483 | 205,774 | 26.71% | 61,856 | 8.03% | 768,242 | 205,233 | 26.71% | 61,854 | 8.05% | 3,347 | 477 | 14.25% | 166 | 4.96% | 2,013 | 218 | 10.83% | 149 | 7.40% | 216,686 | 127,839 | 59.00% | 115,824 | 53.45% |
| Y | 7,110 | 2,239 | 31.49% | 886 | 12.46% | 7,098 | 2,235 | 31.49% | 885 | 12.47% | 20 | 4 | 20.00% | 1 | 5.00% | 6 | 1 | 16.67% | 1 | 16.67% | 10,036 | 7,594 | 75.67% | 3,593 | 35.80% |
| MtDNA | 678 | 344 | 50.74% | 12 | 1.77% | 678 | 326 | 48.08% | 12 | 1.77% | - | - | - | - | - | - | - | - | - | - | 24 | 22 | 91.67% | 17 | 70.83% |

[1]. In order to build a reference genome tailored to the Qatari population, SNP and indel variants were identified in 1,005 unrelated Qataris and both novel and major alternate alleles (MAA) were quantified. A total of 1,161 Qatari DNA samples were sequenced, including 108 selected for genome sequencing and 1,053 selected for exome sequencing. The exome sequencing subjects included n=67 sequenced on the Illumina platform using DNA captured with targets in the Agilent SureSelect Human All Exon 38 Mb platform (Exome38Mb), as well as n=1201 sequenced on the Illumina platform using DNA captured with targets in the Agilent SureSelect Human All Exon 51 Mb platform (Exome51Mb). Variants were identified in each genome or exome using the GATK best practices workflow[8], and the data was integrated using batch-specific filters that ensured a consistent <0.73% novel SNP rate (compared to DbSNP build 146) across platforms in coding intervals covered by all three platforms. Related individuals were identified using variants in the intersection of the three platforms, exclusion of 1st degree and 2nd degree relatives resulted in n=1,005 Qataris, including n=88 genomes and n=917 exomes (n=64 Exome38Mb and n=853 Exome51Mb). The variants frequency was calculated across the n=1,005 Qataris individuals using a Python[43] script, and novel variants were identified using VCFtools[22]. Shown is the total number of SNPs in the integrated set, the number and percentage of novel SNPs, and the number and percentage of MAA SNPs. The same columns are repeated for genome SNPs, Exome50Mb SNPs, Exome38Mb SNPs, and genome indels. The indels were identified using the CASAVA 1.9 workflow[6].

**Supplemental Table VI. Impact of Using QTRG3 on Depth and Variant Sites in Quadruple Sequenced Qatari[1]**

| Platform/batch | GRCh37 Variants | GRCh37 Mean depth | QTRG3 Variants | QTRG3 Mean depth | QTRG3 minus GRCh37 Variants | QTRG3 minus GRCh37 Mean depth |
|---|---|---|---|---|---|---|
| Genome HiSeq 2500 | 369 | 31 | 331 | 32 | -38 | +1 |
| Genome HiSeq X | 369 | 30 | 331 | 32 | -38 | +2 |
| Exome38Mb | 369 | 83 | 331 | 79 | -38 | -4 |
| Exome51Mb | 369 | 90 | 331 | 74 | -38 | -6 |

[1] In order to quantify differences in read depth and variant discovery between QTRG3 and GRCh37 references, sequence data from four platforms (2 genome, 2 exome) was analyzed using GATK Best Practices Workflow[8] twice, once using GRCh37 as the reference and once using QTRG3 as the reference. Shown are the results for Chr20 coding intervals covered in the intersection of the four platforms (Genome HiSeq 2500, Genome HiSeq X, Exome38Mb, Exome51Mb). Calls were filtered to sites covered at least 12* in all four platforms. Shown is the total number of variants observed and mean depth at variant sites for GRCh37, total number of variants observed and mean depth at variant sites for QTRG3, and the difference in total variants and mean depth (negative means lower in QTRG3). Analysis was conducted using VCFtools[22] and Python[43] scripts.

**Supplemental Table VII. Coverage Depth Increase and Variant Call Reduction Using QTRG3 on a Large Qatari Family[1]**

|            | Depth at variant sites |       |          |     | Variants   |           |           |     |           |     |     |
|------------|------------------------|-------|----------|-----|------------|-----------|-----------|-----|-----------|-----|-----|
|            | Observed               |       | Increase |     | Observed   |           | Reduction |     | Expected  |     |     |
| Individual | GRCh37                 | QTRG3 | X        | %   | GRCh37     | QTRG3     | N         | %   | N         | %   | E-O |
| Mean       | 31                     | 38    | 7        | 23% | 4,843,626  | 4,086,955 | 756,671   | 16% | 1,963,417 | 41% | 25% |
| QMD-27-01  | 30                     | 34    | 4        | 14% | 4,713,827  | 3,898,685 | 815,142   | 17% | 1,962,206 | 42% | 24% |
| QMD-27-02  | 26                     | 32    | 6        | 21% | 4,777,169  | 4,017,524 | 759,645   | 16% | 1,956,241 | 41% | 25% |
| QMD-27-04  | 30                     | 37    | 7        | 24% | 4,827,926  | 4,051,563 | 776,363   | 16% | 1,970,911 | 41% | 25% |
| QMD-27-05  | 27                     | 32    | 5        | 19% | 4,841,023  | 4,096,170 | 744,853   | 15% | 1,953,194 | 40% | 25% |
| QMD-27-06  | 30                     | 38    | 8        | 26% | 4,862,510  | 4,100,252 | 762,258   | 16% | 1,968,086 | 40% | 25% |
| QMD-27-07  | 28                     | 35    | 7        | 25% | 4,788,903  | 4,063,026 | 725,877   | 15% | 1,943,632 | 41% | 25% |
| QMD-27-08  | 32                     | 40    | 8        | 24% | 4,891,588  | 4,137,736 | 753,852   | 15% | 1,961,483 | 40% | 25% |
| QMD-27-09  | 30                     | 37    | 7        | 22% | 4,862,523  | 4,106,822 | 755,701   | 16% | 1,970,489 | 41% | 25% |
| QMD-27-10  | 31                     | 38    | 7        | 23% | 4,814,774  | 4,062,059 | 752,715   | 16% | 1,961,938 | 41% | 25% |
| QMD-27-11  | 43                     | 52    | 9        | 22% | 4,949,852  | 4,199,328 | 750,524   | 15% | 1,974,219 | 40% | 25% |
| QMD-27-12  | 30                     | 36    | 6        | 19% | 4,833,472  | 4,078,511 | 754,961   | 16% | 1,969,131 | 41% | 25% |
| QMD-27-13  | 31                     | 38    | 7        | 24% | 4,870,552  | 4,114,454 | 756,098   | 16% | 1,962,077 | 40% | 25% |
| QMD-27-14  | 32                     | 41    | 9        | 27% | 4,888,960  | 4,144,546 | 744,414   | 15% | 1,969,268 | 40% | 25% |
| QMD-27-15  | 31                     | 39    | 8        | 25% | 4,889,792  | 4,137,013 | 752,779   | 15% | 1,967,142 | 40% | 25% |
| QMD-27-16  | 31                     | 39    | 8        | 24% | 4,841,519  | 4,096,639 | 744,880   | 15% | 1,961,233 | 41% | 25% |

[1.] In order to study the impact of a modified reference on genome interpretation, n=15 Qatari genomes from a family of K=9 Persian ancestry were sequenced, and reads were mapped to both GRCh37 and QTRG3 and analyzed using GATK best practices[8]. Genome-wide mapped read depth at variant sites and number of variant sites per genome was quantified using VCFtools[22], limited to sites covered with at least 12* depth. Shown is the mean and individual depth for GRCH37, depth for QTRG3, and the increased depth using QTRG3 (X depth and percentage of GRCh37 depth); the total number of variants in GRCh37 and QTRG3, and the reduction using QTRG3 (number and % of GRCh37); the expected reduction (number of flipped variants) and percentage, and the difference between observed and expected reduction (E-O).

**Supplemental Table VIII. Evaluation of QTRG3 on a Diverse Panel of Qatari Exomes**[1]

| | Sample information | | Variants in Chr20 | | | | | |
|---|---|---|---|---|---|---|---|---|
| | | | Grch37 | | QTRG3 | | | |
| | | | | | Expected | | Observed | |
| ID | Batch | ADMIXTURE Cluster | Variants | MAA | Variants | Reduction (%) | Variants | Reduction (%) |
| DGMQ-31034 | Exome51Mb | K=8 Arab | 1,937 | 758 | 1,179 | 39% | 556 | 71% |
| DGMQ-31094 | Exome51Mb | K=9 Persian | 1,675 | 657 | 1,018 | 39% | 470 | 72% |
| DGMQ-31164 | Exome51Mb | K=5 Bedouin | 1,667 | 674 | 993 | 40% | 470 | 72% |
| DGMQ-31203 | Exome51Mb | K=5 Bedouin | 1,962 | 766 | 1,196 | 39% | 548 | 72% |
| DGMQ-31517 | Exome51Mb | K=4 South Asian | 1,737 | 641 | 1,096 | 37% | 557 | 68% |
| DGMQ-31721 | Exome51Mb | K=5 Bedouin | 2,250 | 878 | 1,372 | 39% | 589 | 74% |
| DGMQ-31728 | Exome51Mb | K=9 Persian | 1,587 | 564 | 1,023 | 36% | 503 | 68% |
| DGMQ-31790 | Exome51Mb | K=9 Persian | 2,204 | 842 | 1,362 | 38% | 587 | 73% |
| DGMQ-32543 | Exome51Mb | K=8 Arab | 1,734 | 733 | 1,001 | 42% | 444 | 74% |
| DGMQ-32513 | Exome51Mb | K=10 Sub-Saharan African | 2,397 | 733 | 1,664 | 31% | 788 | 67% |
| DGMQ-33173 | Exome51Mb | K=10 Sub-Saharan African | 2,164 | 682 | 1,482 | 32% | 640 | 70% |
| DGMQ-33708 | Exome51Mb | K=10 Sub-Saharan African | 2,092 | 636 | 1,456 | 30% | 681 | 67% |
| DGMQ-33147 | Exome51Mb | K=8 Arab | 1,598 | 622 | 976 | 39% | 476 | 70% |
| DGMQ-33553 | Exome51Mb | K=5 Bedouin | 1,866 | 750 | 1,116 | 40% | 450 | 76% |
| DGMQ-33557 | Exome51Mb | K=5 Bedouin | 1,847 | 749 | 1,098 | 41% | 515 | 72% |
| DGMQ-31923 | Exome51Mb | K=8 Arab | 1,851 | 723 | 1,128 | 39% | 549 | 70% |

[1] In order to quantify the impact of using the QTRG3 reference on variant discovery, n=16 exomes from diverse ancestries were mapped to both the GRCh37 and QTRG3 reference genomes, and variants were identified using GATK best practices[8]. Shown is the sample ID, the sequencing batch, the inferred ancestry cluster, the number of variants observed with respect to GRCh37 on Chr20 at, the number of variants that were modified in the reference that are present in this individual, the expected and observed number of variants and percent reduction of variants compared to GRCh37 mapping. The expected variants is equal to the GRCh37 variants minus the flipped in QTRG3 variants, while the observed variants is based on mapping to QTRG3. The reduction is significant with a Chi-square test p value < 0.001.

## Supplemental Table IX. Functional Classification of Variants in Qatar[1]

| Impact | Function | All variants n | All variants % | Major alt alleles (MAA) n | Major alt alleles (MAA) % |
|---|---|---|---|---|---|
| | All | 26,390,578 | 100.00% | 3,827,678 | 100.00% |
| High | Likely pathogenic | 12,721 | 0.05% | 1,622 | 0.04% |
| Moderate | Potentially pathogenic | 164,821 | 0.62% | 8,944 | 0.23% |
| Low | Unlikely pathogenic | 4,357,128 | 16.51% | 621,631 | 16.24% |
| Modifier | Likely benign | 21,855,908 | 82.82% | 3,195,481 | 83.48% |
| High | Frameshift variant | 2,888 | 0.01% | 617.0 | 0.02% |
| | Stop gained | 2,361 | 0.01% | 43.0 | 0.00% |
| | Splice donor variant & intron variant | 2,193 | 0.01% | 268.0 | 0.01% |
| | Splice acceptor variant & intron variant | 1,799 | 0.01% | 289.0 | 0.01% |
| | Protein protein contact | 1,023 | <0.01% | 18.0 | <0.01% |
| | Frameshift variant & stop gained | 421 | <0.01% | 26.0 | <0.01% |
| | Start lost | 374 | <0.01% | 21.0 | <0.01% |
| | Stop lost | 293 | <0.01% | 38.0 | <0.01% |
| | Frameshift variant & splice region variant | 213 | <0.01% | 72.0 | <0.01% |
| | Stop gained & inframe insertion | 101 | <0.01% | 8.0 | <0.01% |
| | Splice donor variant & splice region variant & intron variant & non coding exon variant | 94 | <0.01% | 14.0 | <0.01% |
| | Splice acceptor variant & splice region variant & intron variant | 94 | <0.01% | 37.0 | <0.01% |
| | Splice acceptor variant & splice region variant & intron variant & non coding exon variant | 87 | <0.01% | 18.0 | <0.01% |
| | Stop gained & splice region variant | 80 | <0.01% | 3.0 | <0.01% |
| | Stop gained & disruptive inframe insertion | 70 | <0.01% | 5.0 | <0.01% |
| | Frameshift variant & stop gained & splice region variant | 68 | <0.01% | 12.0 | <0.01% |
| | Splice donor variant & splice region variant & intron variant | 61 | <0.01% | 17.0 | <0.01% |
| | Splice acceptor variant & splice donor variant & intron variant | 55 | <0.01% | 35.0 | <0.01% |
| | Splice acceptor variant & splice donor variant & splice region variant & intron variant & non coding exon variant | 49 | <0.01% | 7.0 | <0.01% |
| | Frameshift variant & stop lost | 38 | <0.01% | 10.0 | <0.01% |
| | Frameshift variant & start lost | 37 | <0.01% | 12.0 | <0.01% |
| | Frameshift variant & splice acceptor variant & splice region variant & intron variant | 28 | <0.01% | 7.0 | <0.01% |
| | Splice donor variant & splice region variant & 5 prime UTR variant & intron variant | 25 | <0.01% | 2.0 | <0.01% |
| | Frameshift variant & splice donor variant & splice region variant & intron variant | 24 | <0.01% | 6.0 | <0.01% |
| | Exon loss variant & splice acceptor variant & splice donor variant & splice region variant & intron variant | 23 | <0.01% | 0.0 | <0.01% |
| | Stop lost & splice region variant | 19 | <0.01% | 4.0 | <0.01% |
| | Transcript ablation | 18 | <0.01% | 3.0 | <0.01% |
| | Stop gained & inframe insertion & splice region variant | 15 | <0.01% | 0.0 | <0.01% |
| | Splice acceptor variant & splice donor variant & splice region variant & intron variant | 14 | <0.01% | 3.0 | <0.01% |
| | Frameshift variant & stop lost & splice region variant | 10 | <0.01% | 0.0 | <0.01% |
| | Start lost & splice region variant | 10 | <0.01% | 1.0 | <0.01% |
| | Splice donor variant & disruptive inframe deletion & splice region variant & intron variant | 9 | <0.01% | 2.0 | <0.01% |

**Supplemental Table IX. Functional Classification of Variants in Qatar[1] (cont., page 2)**

| Impact | Function | All variants | | Major alt alleles (MAA) | |
|---|---|---|---|---|---|
| | | n | % | n | % |
| | Splice donor variant & splice region variant & 3 prime UTR variant & intron variant | 8 | <0.01% | 3.0 | <0.01% |
| | Start lost & inframe deletion | 8 | <0.01% | 3.0 | <0.01% |
| | Splice acceptor variant & splice region variant & 3 prime UTR variant & intron variant | 8 | <0.01% | 5.0 | <0.01% |
| | Stop lost & inframe deletion | 7 | <0.01% | 0.0 | <0.01% |
| | Splice acceptor variant & inframe deletion & splice region variant & intron variant | 6 | <0.01% | 2.0 | <0.01% |
| | Splice donor variant & 5 prime UTR variant & intron variant | 6 | <0.01% | 2.0 | <0.01% |
| | Splice acceptor variant & 5 prime UTR variant & intron variant | 5 | <0.01% | 0.0 | <0.01% |
| | Frameshift variant & splice acceptor variant & splice donor variant & splice region variant & intron variant | 5 | <0.01% | 0.0 | <0.01% |
| | Frameshift variant & start lost & splice region variant | 5 | <0.01% | 2.0 | <0.01% |
| | Splice acceptor variant & splice region variant & 5 prime UTR variant & intron variant | 5 | <0.01% | 0.0 | <0.01% |
| | Stop gained & disruptive inframe insertion & splice region variant | 5 | <0.01% | 0.0 | <0.01% |
| | Start lost & disruptive inframe insertion | 4 | <0.01% | 1.0 | <0.01% |
| | Splice acceptor variant & splice donor variant & 5 prime UTR truncation & exon loss variant & splice region variant & intron variant | 4 | <0.01% | 0.0 | <0.01% |
| | Splice donor variant & 5 prime UTR truncation & exon loss variant & splice region variant & intron variant | 4 | <0.01% | 0.0 | <0.01% |
| | Frameshift variant & missense variant | 3 | <0.01% | 0.0 | <0.01% |
| | Splice acceptor variant & splice donor variant & splice region variant & 3 prime UTR variant & intron variant | 3 | <0.01% | 0.0 | <0.01% |
| | Splice donor variant & 3 prime UTR variant & intron variant | 3 | <0.01% | 0.0 | <0.01% |
| | Exon loss variant & splice acceptor variant & splice donor variant & 3 prime UTR truncation & exon loss variant & splice region variant & intron variant | 3 | <0.01% | 1.0 | <0.01% |
| | Splice acceptor variant & disruptive inframe deletion & splice region variant & intron variant | 3 | <0.01% | 1.0 | <0.01% |
| | Splice acceptor variant & splice donor variant & disruptive inframe deletion & splice region variant & intron variant | 3 | <0.01% | 0.0 | <0.01% |
| | Splice donor variant & inframe deletion & splice region variant & intron variant | 3 | <0.01% | 0.0 | <0.01% |
| | Start lost & inframe insertion | 3 | <0.01% | 1.0 | <0.01% |
| | Stop gained & disruptive inframe deletion | 3 | <0.01% | 0.0 | <0.01% |
| | Frameshift variant & missense variant & splice region variant | 2 | <0.01% | 0.0 | <0.01% |
| | Exon loss variant & splice acceptor variant & splice region variant & intron variant | 2 | <0.01% | 0.0 | <0.01% |
| | Exon loss variant & splice donor variant & splice region variant & intron variant | 2 | <0.01% | 2.0 | <0.01% |
| | Exon loss variant & splice region variant | 2 | <0.01% | 0.0 | <0.01% |
| | Start lost & disruptive inframe deletion | 2 | <0.01% | 0.0 | <0.01% |
| | Stop lost & inframe deletion & splice region variant | 2 | <0.01% | 0.0 | <0.01% |
| | Exon loss variant | 1 | <0.01% | 0.0 | <0.01% |
| | Frameshift variant & splice acceptor variant & missense variant & splice region variant & intron variant | 1 | <0.01% | 0.0 | <0.01% |
| | Frameshift variant & splice donor variant & splice region variant & synonymous variant & intron variant | 1 | <0.01% | 0.0 | <0.01% |
| | Frameshift variant & stop gained & stop retained variant | 1 | <0.01% | 0.0 | <0.01% |
| | Splice acceptor variant & missense variant & inframe insertion & splice region variant & intron variant | 1 | <0.01% | 0.0 | <0.01% |
| | Splice acceptor variant & splice donor variant & inframe deletion & splice region variant & intron variant | 1 | <0.01% | 0.0 | <0.01% |
| | Splice acceptor variant & splice donor variant & splice region variant & 5 prime UTR variant & intron variant | 1 | <0.01% | 0.0 | <0.01% |
| | Stop gained & splice acceptor variant & missense variant & inframe insertion & splice region variant & intron variant | 1 | <0.01% | 1.0 | <0.01% |
| Moderate | Missense variant | 144,405 | 0.55% | 6,039.0 | 0.16% |
| | Sequence feature | 13,953 | 0.05% | 1,950.0 | 0.05% |
| | Missense variant & splice region variant | 3,179 | 0.01% | 138.0 | 0.00% |

**Supplemental Table IX. Functional Classification of Variants in Qatar[1] (cont., page 3)**

| Impact | Function | All variants | | Major alt alleles (MAA) | |
|---|---|---|---|---|---|
| | | n | % | n | % |
| | Disruptive inframe deletion | 1,208 | <0.01% | 292.0 | 0.01% |
| | Disruptive inframe insertion | 764 | <0.01% | 214.0 | 0.01% |
| | Inframe insertion | 662 | <0.01% | 173.0 | <0.01% |
| | Inframe deletion | 562 | <0.01% | 124.0 | <0.01% |
| | Inframe insertion & splice region variant | 48 | <0.01% | 9.0 | <0.01% |
| | Disruptive inframe deletion & splice region variant | 13 | <0.01% | 1.0 | <0.01% |
| | Inframe deletion & splice region variant | 12 | <0.01% | 3.0 | <0.01% |
| | Disruptive inframe insertion & splice region variant | 11 | <0.01% | 0.0 | <0.01% |
| | Missense variant & disruptive inframe insertion | 1 | <0.01% | 0.0 | <0.01% |
| | 5 prime UTR truncation & exon loss variant | 1 | <0.01% | 0.0 | <0.01% |
| | Missense variant & inframe insertion | 1 | <0.01% | 1.0 | <0.01% |
| | Missense variant & inframe insertion & splice region variant | 1 | <0.01% | 0.0 | <0.01% |
| Low | Sequence feature | 4,191,317 | 15.88% | 607,205.0 | 15.86% |
| | Synonymous variant | 108,870 | 0.41% | 6,041.0 | 0.16% |
| | Splice region variant & intron variant | 24,115 | 0.09% | 4,477.0 | 0.12% |
| | 5 prime UTR premature start codon gain variant | 10,353 | 0.04% | 726.0 | 0.02% |
| | TF binding site variant | 10,053 | 0.04% | 1,388.0 | 0.04% |
| | Splice region variant & non coding exon variant | 6,253 | 0.02% | 1,103.0 | 0.03% |
| | Splice region variant | 2,830 | 0.01% | 469.0 | 0.01% |
| | Splice region variant & synonymous variant | 2,551 | 0.01% | 130.0 | <0.01% |
| | TFBS ablation | 636 | <0.01% | 83.0 | <0.01% |
| | Stop retained variant | 109 | <0.01% | 6.0 | <0.01% |
| | Initiator codon variant | 27 | <0.01% | 2.0 | <0.01% |
| | Splice region variant & stop retained variant | 13 | <0.01% | 0.0 | <0.01% |
| | Splice region variant & downstream gene variant | 1 | <0.01% | 1.0 | <0.01% |
| Modifier | Intergenic region | 10,100,190 | 38.27% | 1,493,966.0 | 39.03% |
| | Intron variant | 6,339,378 | 24.02% | 914,967.0 | 23.90% |
| | Upstream gene variant | 2,912,384 | 11.04% | 425,612.0 | 11.12% |
| | Downstream gene variant | 2,132,880 | 8.08% | 311,645.0 | 8.14% |
| | 3 prime UTR variant | 230,874 | 0.87% | 32,038.0 | 0.84% |
| | Non coding exon variant | 74,461 | 0.28% | 9,087.0 | 0.24% |
| | 5 prime UTR variant | 58,106 | 0.22% | 7,413.0 | 0.19% |
| | TF binding site variant | 5,501 | 0.02% | 424.0 | 0.01% |
| | Intragenic variant | 2,134 | 0.01% | 329.0 | 0.01% |

[1] Variants (SNPs and indels) were functionally annotated using SnpEff[29] using ENSEMBL build 75[28] gene models. Shown are the counts of variants, grouped by functional category. The number of variants is shown for all SNPs and for major allele SNPs, and the percentages of the total number of SNPs (top row). Top section lists all SNPs, middle section list subsets of potentially pathogenic and unlikely pathogenic, and third section lists details of each annotated variant category. Categories are organized from top to bottom by decreasing pathogenicity and decreasing n.

**Supplemental Table X. Known Pathogenic Mendelian Mutations Observed in Two or More Qataris[1]**

| Disease(s) | Inheritance model | Reference PMID | Gene | Gene MIM | Chr | Position GRCh37 | Position QTRG3 | rsID | cDNA Change | A.A. Change | # Q - Hom | # Q - Het | Qat-MAF | ExAC-MAF |
|---|---|---|---|---|---|---|---|---|---|---|---|---|---|---|
| Cowden-like Syndrome | Dominant (LO) | 18678321 | SDHB | 185470 | 1 | 17354297 | 17354627 | rs339927012 | c.487T>C | p.Ser163Pro | 0 | 12 | 0.60% | 1.300% |
| Dominant deafness Type 2B | Dominant | 9843210 | GJB3 | 603324 | 1 | 35250910 | 35223785 | rs74315318 | c.547G>A | p.Glu183Lys | 0 | 2 | 0.12% | 0.051% |
| Mcad Deficiency | Dominant (SD) | 9158144 | ACADM | 607008 | 1 | 76199288 | 76198878 | rs121434283 | c.461C>T | p.Thr154Ile | 0 | 7 | 0.41% | 0.001% |
| Diamond-blackfan anemia | Dominant (RP) | 19061985 | RPL5 | 603634 | 1 | 93301840 | 35817017 | rs121434406 | c.418G>A | p.Gly140Ser | 0 | 3 | 1.71% | 0.013% |
| Fundus Flavimaculatus | Dominant (SD) | 9781034 | ABCA4 | 601691 | 1 | 94473287 | 94472343 | rs28938473 | c.5908C>T | p.Leu1970Phe | 0 | 2 | 0.12% | 0.290% |
| Stargardt Disease 1 | Dominant (SD) | 9973280 | ABCA4 | 601691 | 1 | 94473807 | 94472863 | rs1800553 | c.5882G>A | p.Gly1961Glu | 0 | 36 | 1.79% | 0.505% |
| Stargardt Disease 1 | Dominant (SD) | 9054934 | ABCA4 | 601691 | 1 | 94512602 | 94511636 | rs58331765 | c.2791G>A | p.Val931Met | 0 | 2 | 0.11% | 0.049% |
| Erythrocyte Lactate Transporter Deficiency | Dominant (M) | 10590411 | SLC16A1 | 600682 | 1 | 113456602 | 113455493 | rs72552271 | c.1414G>A | p.Gly472Arg | 0 | 4 | 0.23% | 0.015% |
| Thrombophilia due to Factor 5 Leiden | Dominant (RP) | 3110773 | F5 | 612309 | 1 | 169519049 | 169517298 | rs6025 | c.1601A>G | p.Gln534Arg | 0 | 14 | 0.70% | 97.800% |
| Hereditary Prostate Cancer | Dominant (LO) | 11799394 | RNASEL | 180435 | 1 | 182555149 | 182553554 | rs74315364 | c.793G>T | p.Glu265* | 0 | 5 | 0.29% | 0.399% |
| Familial autoinflammatory Syndrome | Dominant | 49161 | NLRP3 | 606416 | 1 | 247587343 | 247585369 | rs121908147 | c.598G>A | p.Val200Met | 0 | 10 | 0.55% | 0.823% |
| Primary Glaucoma | Dominant (LO) | 15342693 | CYP1B1 | 601771 | 2 | 38302291 | 38302637 | rs9282671 | c.241T>A | p.Tyr81Asn | 0 | 3 | 0.19% | 0.345% |
| Atrioventricularseptal Defect | Dominant (RP) | 12632326 | CRELD1 | 607170 | 3 | 9985136 | 9984961 | rs28942091 | c.985C>T | p.Arg329Cys | 0 | 2 | 0.12% | 0.063% |
| Heterotaxy Syndrome | Dominant | 9916847 | ACVR2B | 602730 | 3 | 38518844 | 38518610 | rs121434437 | c.119G>A | p.Arg40His | 0 | 5 | 0.27% | 0.137% |
| Telomere-related Bone Marrow Failure | Dominant (RP) | 15814878 | TERT | 187270 | 5 | 1294397 | 1294427 | rs121918661 | c.604G>A | p.Ala202Thr | 0 | 2 | 0.12% | 0.023% |
| Familial adenomatous polyposis 1 | Dominant (LO) | 11001924 | APC | 611731 | 5 | 112175240 | 112175677 | rs1801166 | c.3949G>C | p.Glu1317Gln | 0 | 32 | 1.59% | 0.413% |
| Tetralogy of Fallot | Dominant (RP) | 16418214 | NKX2-5 | 600584 | 5 | 172662014 | 172662883 | rs28936670 | c.73C>T | p.Arg25Cys | 0 | 11 | 0.56% | 0.371% |
| Tetralogy of Fallot | Dominant (RP) | 11714651 | NKX2-5 | 600584 | 5 | 172662026 | 172662895 | rs104893904 | c.61G>C | p.Glu21Gln | 0 | 4 | 0.24% | 0.078% |
| Thiopurine S-methyltransferase Deficiency | Dominant (SD) | 9931346 | TPMT | 187680 | 6 | 18130993 | 18131172 | rs56161402 | c.644G>A | p.Arg215His | 0 | 4 | 0.26% | 0.204% |
| Hemochromatosis | Dominant | 10194428 | HFE | 613609 | 6 | 26091185 | 26091026 | rs1800730 | c.193A>T | p.Ser65Cys | 0 | 2 | 0.12% | 1.000% |
| Galbladder Disease 1 | Dominant (LO) | 3459187 | ABCB4 | 171060 | 7 | 87082273 | 20682533 | rs58238559 | c.523A>G | p.Thr175Ala | 1 | 9 | 0.64% | 1.100% |
| Prostate Cancer | Dominant | 12244320 | MSR1 | 153622 | 8 | 16012594 | 16012318 | rs41341748 | c.931C>T | p.Arg311* | 0 | 6 | 0.35% | 0.735% |
| Prostate Cancer | Dominant | 12244320 | MSR1 | 153622 | 8 | 16026077 | 16025799 | rs72552387 | c.574G>T | p.Asp192Tyr | 0 | 10 | 0.50% | 0.197% |
| Familial Hyperlipidemia | Dominant (RP) | 8541837 | LPL | 609708 | 8 | 19813529 | 19812860 | rs268 | c.953A>G | p.Asn318Ser | 0 | 4 | 0.23% | 1.300% |
| Multiple Endocrine Neoplasia | Dominant (RP) | 12466368 | RET | 164761 | 10 | 43609990 | 43609141 | rs77711105 | c.1942G>A | p.Val648Ile | 0 | 5 | 0.27% | 0.009% |
| Recessive Visceral Heterotaxy 5 | Dominant (RP) | 9354794 | NODAL | 601265 | 10 | 72195385 | 72194808 | rs104894169 | c.548G>A | p.Arg183Gln | 0 | 2 | 0.12% | 0.012% |
| Dilated Cardiomyopathy | Dominant (LO) | 14662268 | LDB3 | 605906 | 10 | 88446830 | 88445588 | rs121908338 | c.694G>A | p.Asp232Asn | 0 | 4 | 2.27% | 0.454% |
| Cowden-like Syndrome | Dominant (LO) | 18678321 | SDHD | 602690 | 11 | 111958677 | 111957252 | rs11214077 | c.149A>G | p.His50Arg | 0 | 3 | 0.16% | 0.652% |
| Acute Porphyria | Dominant (M) | 8262523 | HMBS | 609806 | 11 | 118963216 | 118961755 | rs118204113 | c.754G>A | p.Ala252Thr | 0 | 8 | 0.43% | - |
| Cirrhosis | Dominant (LO) | 11372009 | KRT8 | 148060 | 12 | 53298582 | 53298312 | rs11554495 | c.268G>T | p.Gly90Cys | 0 | 3 | 0.36% | 0.505% |
| Cirrhosis | Dominant (LO) | 11372009 | KRT8 | 148060 | 12 | 53298606 | 53298336 | rs57749775 | c.244T>C | p.Tyr82His | 0 | 2 | 0.20% | 0.199% |
| Dominant deafness Type 3B | Dominant | 10471490 | GJB6 | 604418 | 13 | 20797606 | 20717395 | rs104894414 | c.14C>T | p.Thr5Met | 0 | 2 | 0.12% | 0.003% |
| Hereditary Nonpolyposis Colon Cancer Type 7 | Dominant (RP) | 12702580 | MLH3 | 604395 | 14 | 75514138 | 75514289 | rs28756990 | c.2221G>T | p.Val741Phe | 0 | 31 | 1.54% | 1.500% |

**Supplemental Table X. All Known Pathogenic Mendelian Mutations Observed in the Qatari Cohort[1] (cont., page 2)**

| Disease(s) | Inheritance model | Reference PMID | Gene | Gene MIM | Chr | Position GRCh37 | Position QTRG3 | rsID | cDNA Change | A.A. Change | # Q - Hom | # Q - Het | Qat-MAF | ExAC-MAF |
|---|---|---|---|---|---|---|---|---|---|---|---|---|---|---|
| Childhood absence epilepsy 5 | Dominant (RP) | 18514161 | GABRB3 | 137192 | 15 | 27018841 | 27019151 | rs25409 | c.31C>T | p.Pro11Ser | 0 | 9 | 0.56% | 0.235% |
| Spherocytosis Type 4 Anemia, Band 3 Tuca-loosa | Dominant | 1378323 | SLC4A1 | 109270 | 17 | 42335888 | 42335820 | rs28931583 | c.980C>G | p.Pro327Arg | 0 | 4 | 0.21% | 0.024% |
| Hypophosphatemic osteoporosis | Dominant (LO) | 18784102 | SLC9A3R1 | 604990 | 17 | 72745313 | 72744916 | rs35910969 | c.328C>G | p.Leu110Val | 0 | 8 | 0.58% | 0.869% |
| Hypogonadotropic hypogonadism | Dominant | 18272894 | KISS1R | 604161 | 19 | 920708 | 920748 | rs121908499 | c.1157G>C | p.Arg386Pro | 0 | 2 | 0.11% | 0.011% |
| Noninsullin-dependent Diabetes Mellitus | Dominant (RP) | 8432414 | INSR | 147670 | 19 | 7125518 | 7125639 | rs1799816 | c.3034G>A | p.Val1012Met | 0 | 37 | 1.84% | 0.898% |
| Cystinuria | Dominant | 11157794 | SLC7A9 | 604144 | 19 | 33334838 | 33334502 | rs121908484 | c.997C>T | p.Arg333Trp | 0 | 2 | 0.13% | 0.009% |
| Cystinuria | Dominant | 11157794 | SLC7A9 | 604144 | 19 | 33353427 | 33353087 | rs79389353 | c.544G>A | p.Ala182Thr | 0 | 3 | 0.19% | 0.274% |
| Primary Open Angle Glaucoma | Dominant (RP) | 19765683 | NTF4 | 162662 | 19 | 49564992 | 49528991 | rs61732310 | c.263C>T | p.Ala88Val | 0 | 17 | 0.90% | 0.469% |
| Dominant Deafness | Dominant (M) | 15015131 | MYH14 | 608568 | 19 | 50747534 | 50747363 | rs119103280 | c.1150G>T | p.Gly384Cys | 0 | 3 | 1.71% | 0.285% |
| Familial Hypertrophic Cardiomyopathy | Dominant (LO) | 15070570 | TNNI3 | 191044 | 19 | 55667607 | 55667367 | rs77615401 | c.244C>T | p.Pro82Ser | 0 | 5 | 0.25% | 0.164% |
| Keratoconus | Dominant (LO) | 15623752 | VSX1 | 605020 | 20 | 25062683 | 25062503 | rs74315436 | c.50T>C | p.Leu17Pro | 0 | 2 | 0.17% | 0.008% |
| Hypertrophic Cardiomyopathy | Dominant (M) | 11733062 | MYLK2 | 606566 | 20 | 30408160 | 30408106 | rs121908108 | c.284C>A | p.Ala95Glu | 0 | 8 | 0.43% | 0.078% |
| Long QT Syndrome | Dominant (M) | 10219239 | KCNE2 | 603796 | 21 | 35742947 | 35743144 | rs74315448 | c.170T>C | p.Ile57Thr | 1 | 32 | 1.81% | 0.088% |
| Becker Muscular Dystrophy | Dominant (M) | 7881286 | DMD | 300377 | X | 31496398 | 31495994 | rs1800279 | c.8762A>G | p.His2921Arg | 1 | 38 | 3.43% | 2.700% |
| Duchenne Muscular Dystrophy | Dominant | 7881286 | DMD | 300377 | X | 31496431 | 31496027 | rs41305353 | c.8729A>T | p.Glu2910Val | 0 | 15 | 1.29% | 2.100% |
| Non-fatal, non-progressive encephalopathy | Dominant (M) | 11238684 | MECP2 | 300005 | X | 153295997 | 153295135 | rs61753971 | c.1318G>A | p.Gly440Ser | 0 | 3 | 0.28% | 0.012% |
| Recessive early-onset Parkinson Disease | Recessive | 12953260 | PARK7 | 602533 | 1 | 8044990 | 8045123 | rs74315352 | c.446A>C | p.Asp149Ala | 0 | 3 | 0.16% | 0.020% |
| Masp2 Deficiency | Recessive | 12904520 | MASP2 | 605102 | 1 | 11106666 | 11106808 | rs72550870 | c.359A>G | p.Asp120Gly | 0 | 3 | 0.16% | 2.200% |
| Limb-girdle muscular dystrophy | Recessive | 18195152 | POMGNT1 | 606822 | 1 | 46655645 | 46656025 | rs74374973 | c.1666G>A | p.Asp556Asn | 0 | 6 | 0.33% | 0.893% |
| Cystathioniuria | Recessive | 12574942 | CTH | 607657 | 1 | 70881670 | 70777863 | rs28941785 | c.200C>T | p.Thr67Ile | 0 | 3 | 0.16% | 0.647% |
| Myopathy due to Myoadenylate Deaminase Deficiency | Recessive | 10996775 | AMPD1 | 102770 | 1 | 115220593 | 110169903 | rs35859650 | c.1261C>T | p.Arg421Trp | 0 | 4 | 0.21% | 0.068% |
| Gaucher Disease type I | Recessive | 22713811 | GBA | 606463 | 1 | 155205043 | 155183055 | rs421016 | c.1448T>C | p.Leu483Pro | 0 | 3 | 0.18% | 0.316% |
| Recessive spherocytosis Type 3 | Recessive | 3785322 | SPTA1 | 182860 | 1 | 158624528 | 158623150 | rs35948326 | c.2909C>A | p.Ala970Asp | 0 | 11 | 0.64% | 3.300% |
| Porphyria | Recessive | 11286631 | PPOX | 600923 | 1 | 161138933 | 161137548 | rs12735723 | c.767C>G | p.Pro256Arg | 0 | 2 | 0.12% | 0.665% |
| Nephrotic Syndrome | Recessive | 12464671 | NPHS2 | 604766 | 1 | 179526214 | 179524646 | rs61747728 | c.686G>A | p.Arg229Gln | 0 | 18 | 0.98% | 2.900% |
| Nephrotic Syndrome | Recessive | 10742096 | NPHS2 | 604766 | 1 | 179544941 | 179543383 | rs74315344 | c.59C>T | p.Pro20Leu | 0 | 6 | 0.40% | 0.195% |
| Primary Glaucoma | Recessive (RP) | 10655546 | CYP1B1 | 601771 | 2 | 38298394 | 38298741 | rs79204362 | c.1103G>A | p.Arg368His | 0 | 48 | 2.39% | 0.596% |
| Rod Monochromacy (colorblindness) | Recessive | 9662398 | CNGA3 | 600053 | 2 | 99012480 | 99012626 | rs104893613 | c.859C>T | p.Arg287Trp | 0 | 3 | 0.18% | 0.014% |
| Recessive Thrombophilia due to Protein C Deficiency | Recessive | 1678832 | PROC | 612283 | 2 | 128186038 | 128185850 | rs121918144 | c.1067C>T | p.Ala356Val | 0 | 7 | 0.41% | - |
| Odontoonychodermal Dysplasia (Tooth agenesis) | Recessive | 19559398 | WNT10A | 606268 | 2 | 219755011 | 219752068 | rs121908120 | c.682T>A | p.Phe228Ile | 0 | 9 | 0.50% | 1.300% |
| Long QT Syndrome | Recessive | 17060380 | CAV3 | 601253 | 3 | 8787330 | 8787143 | rs72546668 | c.233C>T | p.Thr78Met | 0 | 20 | 1.01% | 0.303% |
| Biotinidase Defeicency | Recessive | 10400129 | BTD | 609019 | 3 | 15686693 | 15686407 | rs13078881 | c.1336G>C | p.Asp446His | 0 | 35 | 1.74% | 3.200% |
| Hurler Syndrome | Recessive | 8328452 | IDUA | 252800 | 4 | 996555 | 996554 | rs11934801 | c.1291G>C | p.Gly431Arg | 0 | 3 | 1.71% | 0.388% |

Supplemental Table X. All Known Pathogenic Mendelian Mutations Observed in the Qatari Cohort[1] (cont., page 3)

| Disease(s) | Inheritance model | Reference PMID | Gene | Gene MIM | Chr | Position GRCh37 | Position QTRG3 | rsID | cDNA Change | A.A. Change | # Q - Hom | # Q - Het | Qat-MAF | ExAC-MAF |
|---|---|---|---|---|---|---|---|---|---|---|---|---|---|---|
| Hypogonadotropic hypogonadism | Recessive | 9371856 | GNRHR | 138850 | 4 | 68619737 | 68618073 | rs104893836 | c.317A>G | p.Gln106Arg | 0 | 2 | 0.12% | 0.252% |
| Recessive Dyskeratosis Congenita | Recessive | 15814878 | TERT | 187270 | 5 | 1293767 | 1293797 | rs34094720 | c.1234C>T | p.His412Tyr | 1 | 3 | 0.29% | 0.224% |
| Recessive Neuropathy | Recessive | 16399879 | CCT5 | 610150 | 5 | 10256175 | 10256372 | rs118203986 | c.440A>G | p.His147Arg | 1 | 8 | 0.53% | 0.004% |
| Central hypoventilation Syndrome | Recessive | 9497256 | GDNF | 600837 | 5 | 37816112 | 37816039 | rs36119840 | c.328C>T | p.Arg110Trp | 0 | 2 | 0.11% | 0.221% |
| C7 Deficiency | Recessive | 9218625 | C7 | 217070 | 5 | 40955530 | 40955516 | rs121964921 | c.1135G>C | p.Gly379Arg | 0 | 2 | 1.14% | 0.013% |
| Laron Dwarfism | Recessive (M) | 9814495 | GHR | 600946 | 5 | 42699970 | 42699907 | rs6413484 | c.484G>A | p.Val162Ile | 0 | 3 | 0.18% | 0.137% |
| Laron Dwarfism | Recessive (M) | 7565946 | GHR | 600946 | 5 | 42700021 | 42699958 | rs121909362 | c.535C>T | p.Arg179Cys | 0 | 2 | 0.12% | 0.409% |
| Progressive Arthropathy | Recessive | 10471507 | WISP3 | 603400 | 6 | 112382323 | 112381757 | rs17073260 | c.232C>T | p.Arg78Cys | 1 | 11 | 0.65% | 0.937% |
| Recessive Parkinson disease 2 (juvenile) | Recessive | 11487568 | PARK2 | 602544 | 6 | 162683724 | 162683679 | rs55774500 | c.245C>A | p.Ala82Glu | 0 | 4 | 0.21% | 0.474% |
| Maple syrup Urine Disease | Recessive | 9934985 | DLD | 238331 | 7 | 107555951 | 107555906 | rs121964990 | c.685G>T | p.Gly229Cys | 0 | 4 | 0.23% | 0.026% |
| Congenital Bilateral Absense of the Vas Deferens | Recessive | 1545465 | CFTR | 602421 | 7 | 117230454 | 117230209 | rs1800098 | c.1727G>C | p.Gly576Ala | 0 | 4 | 0.24% | 0.513% |
| Cystic Fibrosis | Recessive | 7504969 | CFTR | 602421 | 7 | 117267807 | 117267564 | rs75389940 | c.3700A>G | p.Ile1234Val | 0 | 9 | 0.53% | 0.001% |
| Recessive myotonia | Recessive | 11113225 | CLCN1 | 118425 | 7 | 143048886 | 143048936 | rs80356706 | c.2795C>T | p.Pro932Leu | 0 | 2 | 0.12% | 0.014% |
| Retinitis pigmentosa | Recessive | 15863674 | RP1 | 603937 | 8 | 55537560 | 55536834 | rs77775126 | c.1118C>T | p.Thr373Ile | 2 | 34 | 1.89% | 1.300% |
| Recessive myopathy | Recessive | 11528398 | GNE | 603824 | 9 | 36217396 | 36217255 | rs28937594 | c.2228T>C | p.Met743Thr | 0 | 2 | 0.12% | 0.003% |
| Aldolase B Deficiency | Recessive | 3383242 | ALDOB | 612724 | 9 | 104189856 | 104188749 | rs1800546 | c.448G>C | p.Ala150Pro | 0 | 3 | 0.18% | 0.270% |
| Cockayne Syndrome Type B | Recessive | 9443879 | ERCC6 | 609413 | 10 | 50678722 | 50678022 | rs4253208 | c.3284C>G | p.Pro1095Arg | 0 | 4 | 0.23% | 0.378% |
| Hemophagocytic Lymphohistiocytosis (including non-hodgkin lymphoma) | Recessive (RP) | 12229880 | PRF1 | 170280 | 10 | 72360387 | 72359805 | rs35947132 | c.272C>T | p.Ala91Val | 0 | 9 | 0.50% | 3.100% |
| Dubin-Johnson Syndrome | Recessive | 11477083 | ABCC2 | 601107 | 10 | 101595950 | 101594812 | rs72558201 | c.3517A>T | p.Ile1173Phe | 0 | 2 | 0.12% | 0.002% |
| Beta-Thalassemia | Recessive (M; Mod) | 16114182 | HBB | 141900 | 11 | 5248173 | 5248158 | rs33950507 | c.79G>A | p.Glu27Lys | 0 | 4 | 0.21% | 0.029% |
| Sickle Cell Anemia | Recessive | 13852872 | HBB | 141900 | 11 | 5248232 | 5248217 | rs334 | c.20A>T | p.Glu7Val | 0 | 3 | 1.71% | 0.439% |
| Beta-Thalassemia | Recessive (M; Mod) | 2467892 | HBD | 142000 | 11 | 5255582 | 5248155 | rs35152987 | c.82G>T | p.Ala28Ser | 0 | 7 | 0.41% | 0.223% |
| Congenital Central Hypoventilation Syndrome | Recessive | 11840487 | BDNF | 113505 | 11 | 27680107 | 27678988 | rs8192466 | c.251C>T | p.Thr84Ile | 0 | 24 | 1.19% | 0.109% |
| Omenn Syndrome | Recessive | 9630231 | RAG1 | 179615 | 11 | 36596041 | 36594713 | rs104894291 | c.1187G>A | p.Arg396His | 0 | 2 | 0.12% | 0.002% |
| Oculocutaneous Albinism Type IA | Recessive | 1642278 | TYR | 606933 | 11 | 88911770 | 88910230 | rs63159160 | c.649C>T | p.Arg217Trp | 0 | 5 | 0.25% | 0.020% |
| Oculocutaneous Albinism Type IB | Recessive | 5516239 | TYR | 606933 | 11 | 89017973 | 49435662 | rs104894313 | c.1217C>T | p.Pro406Leu | 1 | 3 | 0.27% | 0.249% |
| Hyperalphalipoproteinemia 2 | Recessive | 24941082 | APOC3 | 107720 | 11 | 116701353 | 116699867 | rs76353203 | c.109C>T | p.Arg37* | 0 | 2 | 1.14% | 0.068% |
| Recessive Bartter Syndrome | Recessive | 8841184 | KCNJ1 | 600359 | 11 | 128709126 | 17408421 | rs59172778 | c.1070T>C | p.Met357Thr | 0 | 6 | 0.32% | 0.802% |
| Adrenoleukodystrophy | Recessive | 7719337 | PEX5 | 600414 | 12 | 7362296 | 7362469 | rs61752138 | c.1641T>G | p.Asn547Lys | 0 | 2 | 0.12% | - |
| Recessive Nephrogenic Diabetes Insipidus | Recessive | 9048343 | AQP2 | 107777 | 12 | 50344816 | 50344595 | rs104894331 | c.203A>C | p.Asn68Thr | 0 | 24 | 1.33% | 4.700% |
| Vitamin D Hydroxylation-Defiiency Rickets | Recessive | 12050193 | CYP27B1 | 609506 | 12 | 58159103 | 58158702 | rs118204012 | c.566A>G | p.Glu189Gly | 0 | 8 | 0.45% | 0.121% |
| Hyper-IgD Syndrome | Recessive | 10369261 | MVK | 251170 | 12 | 110034320 | 110034227 | rs28934897 | c.1129G>A | p.Val377Ile | 0 | 6 | 0.32% | 0.140% |

Supplemental Table X. All Known Pathogenic Mendelian Mutations Observed in the Qatari Cohort[1] (cont., page 4)

| Disease(s) | Inheritance model | Reference PMID | Gene | Gene MIM | Chr | Position GRCh37 | Position QTRG3 | rsID | cDNA Change | A.A. Change | # Q - Hom | # Q - Het | Qat-MAF | ExAC-MAF |
|---|---|---|---|---|---|---|---|---|---|---|---|---|---|---|
| Peeling skin Syndrome | Recessive | 16380904 | TGM5 | 603805 | 15 | 43552349 | 43552928 | rs112292549 | c.337G>T | p.Gly113Cys | 0 | 3 | 0.18% | 0.259% |
| Tay-sachs Disease | Recessive | 2140574 | HEXA | 606869 | 15 | 72637817 | 72638091 | rs121907956 | c.1529G>A | p.Arg510His | 0 | 2 | 0.12% | 0.007% |
| Fumarylacetoacetase Deficiency | Recessive | 7977370 | FAH | 613871 | 15 | 80472526 | 80472708 | rs11555096 | c.1021C>T | p.Arg341Trp | 1 | 9 | 0.55% | 1.700% |
| Familial Mediterranean Fever | Recessive | 9668175 | MEFV | 608107 | 16 | 3293403 | 3293353 | rs104895094 | c.2084A>G | p.Lys695Arg | 0 | 3 | 0.18% | 0.550% |
| Bardet-Biedl Syndrome 2 | Recessive (Digenic) | 11567139 | BBS2 | 606151 | 16 | 56548501 | 56548300 | rs4784677 | c.209G>A | p.Asn70Ser | 1 | 10 | 0.60% | 0.600% |
| Congenital Myasthenic syndrome | Recessive | 8755487 | CHRNE | 100725 | 17 | 4805239 | 4805392 | rs121909516 | c.488C>G | p.Ser163Trp | 0 | 4 | 0.24% | 0.003% |
| Congenital ichthyosiform Erythrodermia | Recessive | 16116617 | ALOXE3 | 607206 | 17 | 8015486 | 7946150 | rs121434235 | c.1177T>A | p.Leu393Met | 0 | 12 | 0.70% | 0.558% |
| Smith-Magenic Syndrome | Recessive | 11735029 | MYO15A | 602666 | 17 | 18051447 | 18051368 | rs121908970 | c.6614C>T | p.Thr2205Ile | 0 | 2 | 0.12% | 0.417% |
| Myeloperoxidase Deficiency | Recessive | 9354683 | MPO | 606989 | 17 | 56356502 | 56272604 | rs56378716 | c.848T>C | p.Met283Thr | 0 | 12 | 0.64% | 1.000% |
| Erythropoietic protoporphyria | Recessive | 1755842 | FECH | 612386 | 18 | 55226380 | 55226136 | rs118204037 | c.819G>A | p.Met273Ile | 0 | 17 | 0.85% | 0.109% |
| Hypogonadotropic hypogonadism | Recessive | 17164310 | KISS1R | 604161 | 19 | 918604 | 918644 | rs104894703 | c.305T>C | p.Leu102Pro | 0 | 5 | 0.30% | 0.003% |
| Dominant hyperlipoprotenemia Type 3 | Recessive | 9649566 | APOE | 107741 | 19 | 45412040 | 45411899 | rs769455 | c.487C>T | p.Arg163Cys | 0 | 4 | 0.35% | 0.123% |
| Limb-girdle muscular dystrophy | Recessive | 14647208 | FKRP | 606596 | 19 | 47258942 | 47258803 | rs104894683 | c.235G>A | p.Val79Met | 0 | 6 | 0.35% | 0.117% |
| Kallmann Syndrome | Recessive | 17054399 | PROKR2 | 607123 | 20 | 5294762 | 5294846 | rs74315418 | c.254G>A | p.Arg85His | 0 | 3 | 0.18% | 0.074% |
| Adenomise Deaminase Deficiency | Recessive | 2166947 | ADA | 608958 | 20 | 43255233 | 43255235 | rs121908736 | c.226C>T | p.Arg76Trp | 0 | 2 | 0.11% | 0.035% |
| Arterial Tortuosity Syndrome | Recessive | 18565096 | SLC2A10 | 606145 | 20 | 45353918 | 45353942 | rs80358230 | c.243C>G | p.Ser81Arg | 2 | 15 | 0.95% | - |
| Homocystinuria | Recessive | 10364517 | CBS | 613381 | 21 | 44483184 | 44483437 | rs5742905 | c.833T>C | p.Ile278Thr | 0 | 98 | 4.89% | 0.007% |
| Leukocyte adhesion Deficiency | Recessive | 1346613 | ITGB2 | 600065 | 21 | 46309312 | 46309674 | rs5030672 | c.1756C>T | p.Arg586Trp | 0 | 3 | 0.18% | 0.755% |
| Hyperprolinemia Type I | Recessive | 17412540 | PRODH | 606810 | 22 | 18905859 | 18905888 | rs2870984 | c.1397C>T | p.Thr466Met | 0 | 15 | 0.99% | 0.573% |
| Schizophrenia and mild hyperprolinemia | Recessive | 12217952 | PRODH | 606810 | 22 | 18905899 | 18905928 | rs3970559 | c.1357C>T | p.Arg453Cys | 0 | 19 | 1.37% | 1.200% |
| Schizophrenia and mild hyperprolinemia | Recessive | 12217952 | PRODH | 606810 | 22 | 18905934 | 18905963 | rs2904551 | c.1322T>C | p.Leu441Pro | 0 | 6 | 0.51% | 0.575% |
| Glucose-6-Phosphate Dehydrogenase Deficiency | Recessive (M) | 3393536 | G6PD | 305900 | X | 153761205 | 153760352 | rs5030869 | c.1141G>A | p.Ala381Thr | 0 | 2 | 0.20% | 0.023% |
| Glucose-6-Phosphate Dehydrogenase Deficiency | Recessive (M; Mod) | 3393536 | G6PD | 305900 | X | 153762634 | 153761781 | rs5030868 | c.563C>T | p.Ser188Phe | 1 | 47 | 4.20% | 0.310% |
| Glucose-6-Phosphate Dehydrogenase Deficiency | Recessive | 3393536 | G6PD | 305900 | X | 153764371 | 153763518 | rs76645461 | c.143T>C | p.Ile48Thr | 0 | 4 | 0.37% | 0.001% |

[1] Curated list of known pathogenic Mendelian Disease causing mutations observed in at least 2 Qataris. Table is divided by inheritance patterns (dominant, recessive) and sorted by chromosome and position. Reported for each mutation is the disease it causes, the inheritance model, Pubmed ID (PMID) for the reference publication where the causality was initially reported, gene name, gene MIM number, variant site in GRCh37 (chromosome, position, rsID) and the effect of the mutation on cDNA and on amino acid. Also reported is the total number of homozygous and heterozygous Qataris observed harboring this mutation (# Q - Hom & Het, respectively), the overall frequency of the deleterious allele in Qataris, and allele frequency in the Exome Aggregation Consortium (ExAC, total alleles ~63,000). LO: Late Onset, SD: Semi-dominant, M: Mild, RP: Reduced Penetrance. PMID: Pubmed ID. MAF: Minor Allele Frequency. Positions include both GRCh37 and QTRG3 coordinates inferred using HALtools[44] liftover of GRCh37 to QTRG3 alignment produced using progressiveCactus[26].

## Supplemental Figure Legends

**Supplemental Figure 1.** Sample analysis flowchart. Shown is an overview of the major analysis steps and samples included in each analysis. From top to bottom and left to right, samples for sequencing in this study include n=215 s living in Qatar and n=1161 Qataris, defined as such based on 3 generations of ancestry in Qatar. Exome sequencing was conducted on n=1268 samples using Illumina instruments and two Agilent SureSelect Human All Exon kits (n=1201 and Qatari on the 51 Mb kit and n=67 Qatari on the 38 Mb kit), hereafter referred to as "Exome38Mb" and "Exome51Mb". Genome sequencing was also conducted for n=108 Qatari samples on an Illumina instrument. Variant calling was conducted for each batch of samples (Exome51Mb, Exome38Mb, genome), and combined into a preliminary SNP and indel call set. Using variant calls within genomic intervals covered in all three batches (within CCDS coding exons targeted by the Agilent SureSelect 38 Mb kit), three preliminary analyses were conducted. The preliminary analysis included (1) identification of optimal batch-specific filters, (2) identification of relatives in the n=1376 samples using a LD-pruned set of SNPs, and (3) population structure analysis (ADMIXTURE and principal components) using a subset of variants also segregating in the database of public samples (Human Origins and 1000 Genomes). Based on the results of preliminary analysis, pruning of non-Qataris (n=215) and $1^{st}/2^{nd}$ degree relatives (n=156) resulted in n=1005 unrelated Qataris for further analysis, including n=917 exomes (n=64 Exome38Mb and n=853 Exome51Mb), and n=88 genomes. The allele frequencies of the remaining variants was quantified, and a comparison to dbSNP 146 was conducted to identify novel variants and enable database annotation of known variants. Using variants (SNPs and indels) present in over 50% of the n=1005 samples, the GRCh37 human reference genome was modified to incorporate these major alternate alleles (MAA). The value of using the modified reference, or Qatar Reference Genome (QTRG) was then assessed by comparing sequencing depth and variant calls for Qataris

not in the n=1005 used to construct the reference. In addition, a catalog of potentially pathogenic variants was constructed by combining the Qatari variant allele frequency information with public database for genes and variants linked to disease and algorithms for predicting variant function and pathogenicity.

**Supplemental Figure 2**. Impact of minimum depth and minimum allele count on novel SNP rate. A total of 1376 samples were sequenced using Illumina technology on three platforms, including genome (n=108), exome sequencing based on enrichment using the Agilent SureSelect 38 Mb targets (Exome38Mb, n=67), and exome sequencing based on enrichment using the Agilent SureSelect 51 Mb targets (Exome51Mb, n=1201). In order to integrate the three datasets, genotypes were generated for a set of variants covered by the three platforms, including CCDS coding exons overlapping the Exome38Mb targets. Variants were filtered using a range of minimum depth and minimum allele count parameters, shown is the impact of these filters on **A.** Mean depth at variant sites, **B.** Mean number of variant sites per sample, and **C.** percentage of novel variants, defined as not present in DbSNP build 146. Dotted line indicates level of consistent novel SNP rate.

**Supplemental Figure 3.** Impact of filters on batch-specific variants. Batch-specific variants. In order to validate the selected minimum depth and minimum allele count for each platform, the proportion of batch-specific variants was compared in variant call data with and without filters applied. Data source is a single Qatari female DNA sample that was sequenced on four platforms (2 exome, 2 genome), including genome sequencing on both the Illumina HiSeq 2500 and Illumina HiSeq X platforms, as well as Illumina sequencing of exome DNA enriched using the Agilent SureSelect Human All Exon 51 Mb and 38 Mb targets. Variant calls were generated using GATK HaplotypeCaller on the four datasets, limited to CCDS coding intervals covered by the four platforms (within 38 Mb targets). Shown are Venn diagrams of **A.** unfiltered variant sites,

**B.** unfiltered novel variant sites (not in DbSNP 146), **C.** filtered variant sites, and **D.** filtered novel variant sites.

**Supplemental Figure 4.** Impact of filters on sensitivity. In order to determine the impact of batch-specific filters on sensitivity for variant detection and novel SNP rate across a range of depths, variant calls were generated for a quadruple sequenced Qatari using the GATK "dfrac" parameter, which enables variant calling on a subsampling a percentage of reads. The dfrac-based calling was conducted for a range of values from 5% to 100%, in 5% increments. At each setting, calls were generated on genomic intervals overlapping between the four platforms four platforms (2 exome, 2 genome), including genome sequencing on both the Illumina HiSeq 2500 and Illumina HiSeq X platforms, as well as Illumina sequencing of exome DNA enriched using the Agilent SureSelect Human All Exon 51 Mb and 38 Mb targets. Variant calls were generated using GATK HaplotypeCaller on the four datasets, limited to CCDS coding intervals covered by the four platforms (within 38 Mb targets). Shown is **A.** the total number of unfiltered variants (y-axis) in the four platforms across a range of depths (x-axis) color-coded by platform (blue = Exome38Mb, green = Exome51Mb, red = genome on HiSeq 2500, orange = genome on HiSeq X). **B.** total number of filtered variants (y-axis) in the four platforms across a range of depths (x-axis), **C.** the proportion of unfiltered novel variants (y-axis) in the four platforms across a range of depths (x-axis), and **D.** the proportion of filtered novel variants (y-axis) in the four platforms across a range of depths (x-axis).

**Supplemental Figure 5.** Dominant ancestries. In order to infer the dominant ancestry of each of the n=1161 Qatari samples included in this study, and ADMIXTURE analysis was conducted on n=5611 samples, including the n=1161 Qataris included in this study, the n=215 s sampled in Qatar, the Human Origins data (n=1862), and the 1000 Genomes Phase 3 data (n=2373 not in Human Origins). Integration of the datasets resulted in n=2256 SNPs present in all three (n=1376

from this study, Human Origins, and 1000 Genomes). The ADMIXTURE analysis was conducted with K=12, determined to be optimal in a prior study[17], resulting in admixture proportions in 12 ancestral populations for each individual. Individuals were assigned to an ancestral population based on their dominant ancestry, and cluster names were assigned based on the major populations present in these clusters. Shown is the ancestry proportion in **A.** the full dataset (n=5611), and a detailed view of the results from the same analysis in **B.** the n=1161 Qatari.

**Supplemental Figure 6.** Principal components analysis. In order to confirm the observed population structure inferred using ADMIXTURE[16,26], a principal components (PC) analysis[45] was conducted on the n=5611 samples, including n=1161 Qataris, n=215 s living in Qatar, n=1862 Human Origins samples, and n=2373 1000 Genomes Phase 3 samples not in Human Origins. Analysis was conducted using the same n=2256 SNPs used in the ADMIXTURE[16,26] analysis. Each individual is color-coded based on their dominant ancestry (K cluster), with names assigned based on the major populations in the cluster (K=1 European, K=2 East Asian, K=3 Siberian, K=4 South Asian, K=5 Bedouin, K=6 African Pygmy, K=7 Oceanian, K=8 Arab, K=9 Persian, K=10 Sub-Saharan African, K=11 American, K=12 Central Asian). Shown is a plot of **A.** PC1 *vs* PC2 and **B.** PC2 *vs* PC3.

**Supplemental Figure 7.** Relatives in Qatar. In order to compare inferred pedigree sizes between Qataris and non-Qataris, a graph of inferred relatives was produced. Shown are inferred pedigrees for **A.** Qataris and **B.** s, in order of decreasing size. Relationships were inferred between n=1376 samples (including Qataris and s) as described in the Supplemental Methods and Summarized in Supplemental Table IV. Each node represents and individual, and the line connecting two individuals represents the relationship (1st degree = straight line, 2nd degree = wavy line, 3rd degree = dotted line). The nodes are color-coded based on the inferred dominant ancestry in the ADMIXTURE[16,26] analysis.

## Online Resources

**Online Resource 1. Prevalence and Function of Qatari Variants.** Tab-delimited file including allele frequency in the Qatari population, number of homozygous and heterozygous individuals, functional annotation, and links to databases of gene and variant function for 20.9 million SNPs identified in 88 unrelated Qatari genomes 917 unrelated Qatari exomes. Available for download is the full list of variants, as well as subsets of category 1, category 2, and category 3 variants, as described in Table II. Variant database can be accessed at http://geneticmedicine.weill.cornell.edu/Genome/Online.Resource.1/

**Online Resource 2. Qatari Reference Genome (QTRG1, QTRG2, QTRG3).** Human reference genome based on GRCh37 where the major alternate allele identified in 88 unrelated Qatari genomes 917 unrelated Qatari exomes replaced the reference allele SNP and indel sites. Accessible at http://geneticmedicine.weill.cornell.edu/Genome/Online.Resource.2/.

**Online Resource 3. Precision medicine tools for Qatari genomes.** Genome interpretation analysis workflow based on GATK Best Practices[8] for an individual Qatari genome, coded in Python. Accessible at http://geneticmedicine.weill.cornell.edu/Genome/Online.Resource.3/.

**Online Resource 4. All variants observed in unrelated Qataris.** Genotypes for variant sites (with respect to GRCh37) observed in 1376 Qataris, in VCF format before application of platform specific filters. Accessible at http://geneticmedicine.weill.cornell.edu/Genome/Online.Resource.4/.

**Online Resource 5. Segregating sites in unrelated Qataris.** Genotypes for variant sites polymorphic in 1005 unrelated Qataris, in PLINK format[12]. Accessible at http://geneticmedicine.weill.cornell.edu/Genome/Online.Resource.5/.

# Supplemental References

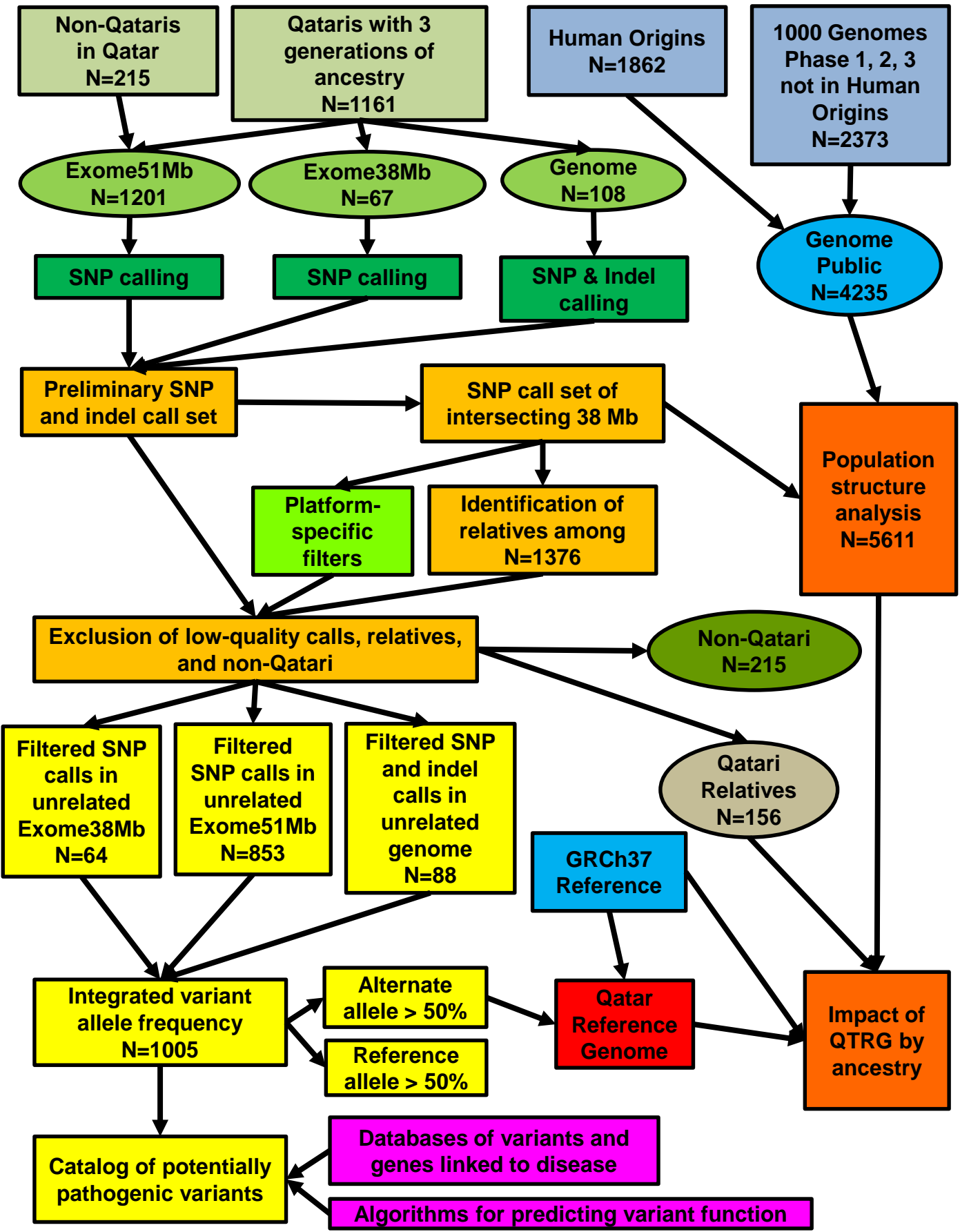

**Supplemental Figure 2**

## A. Mean depth

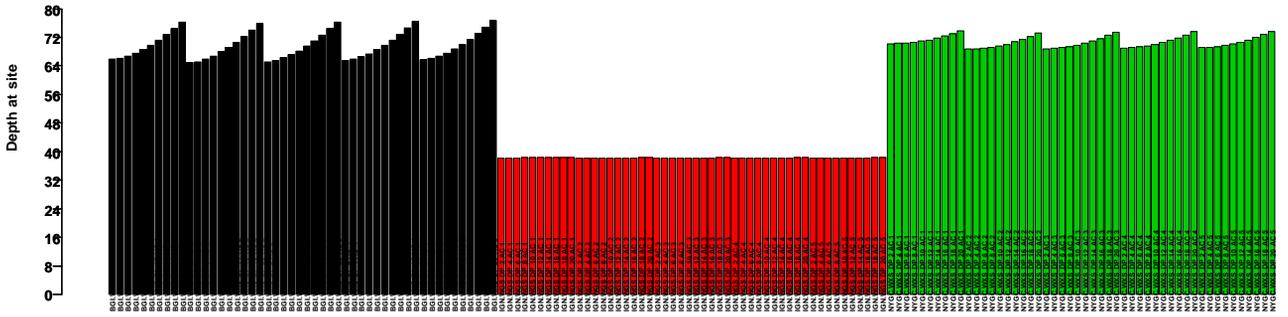

Mean, ordered by platform, minimum allele count (AC), minimum depth (DP)

## B. Variant sites

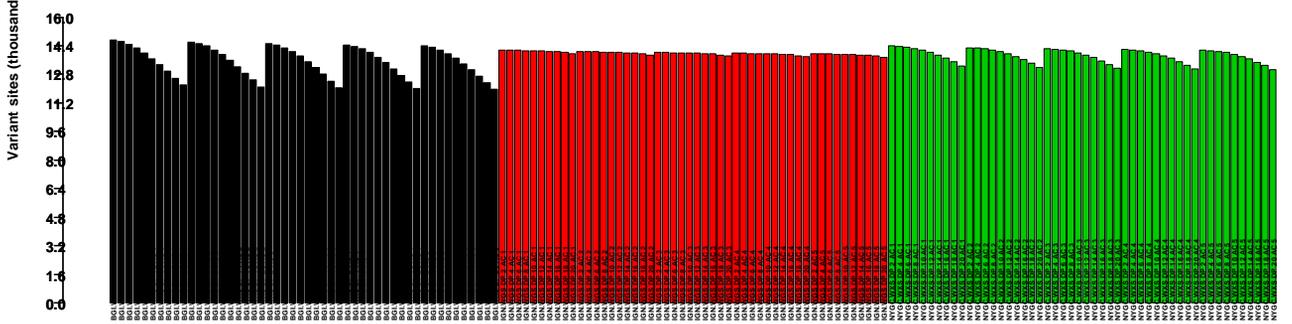

Mean, ordered by platform, minimum allele count (AC), minimum depth (DP)

## C. Novel SNPs (% of variants)

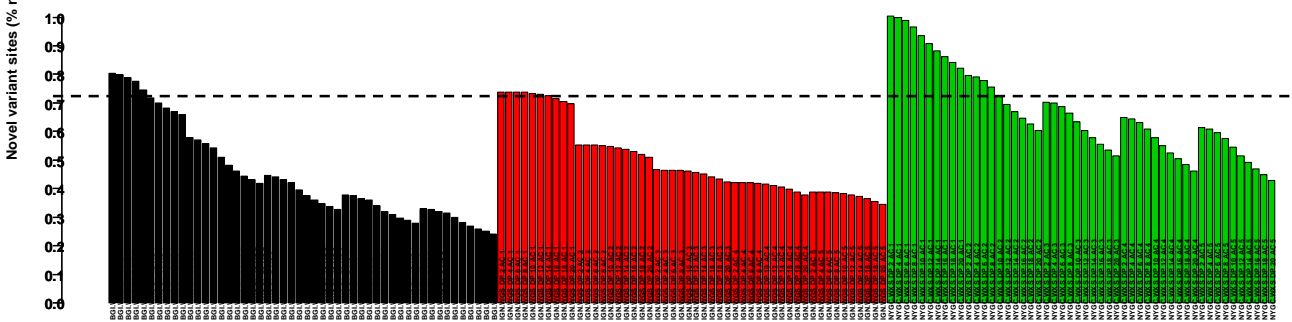

Mean, ordered by platform, minimum allele count (AC), minimum depth (DP)

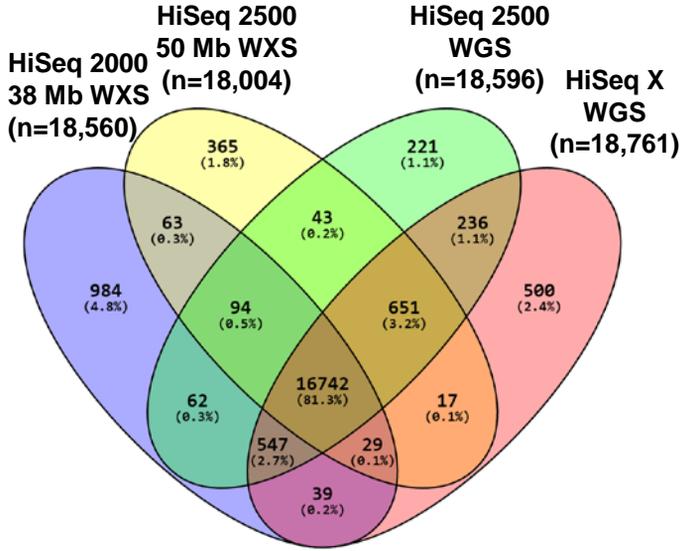

A. Variants (unfiltered)

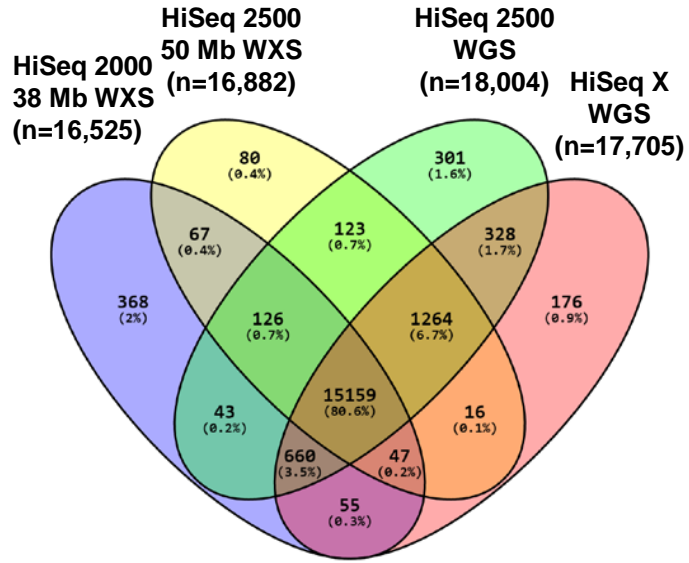

C. Variants (platform-specific filters applied)

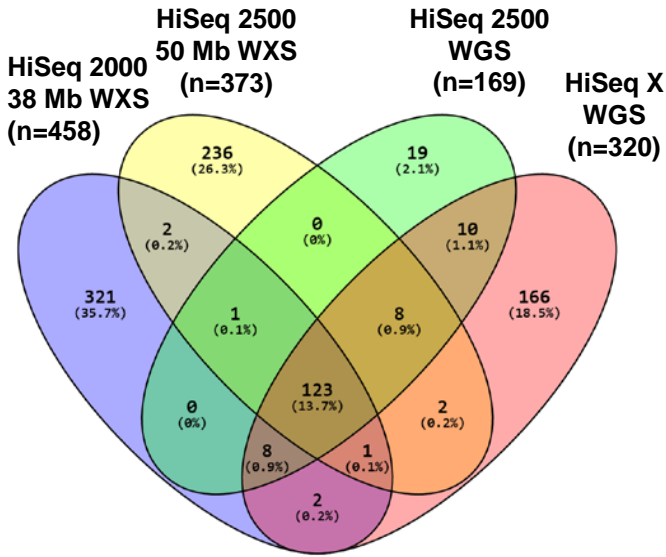

B. Novel variants (unfiltered)

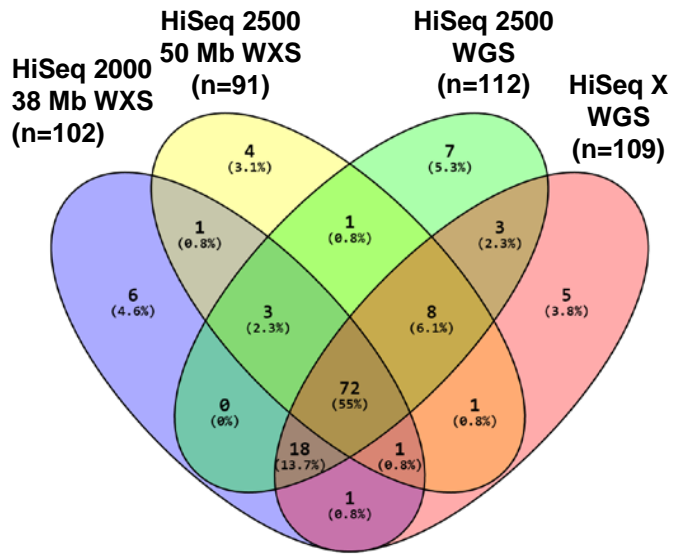

D. Novel variants (platform-specific filters applied)

**A. Variants (unfiltered)**

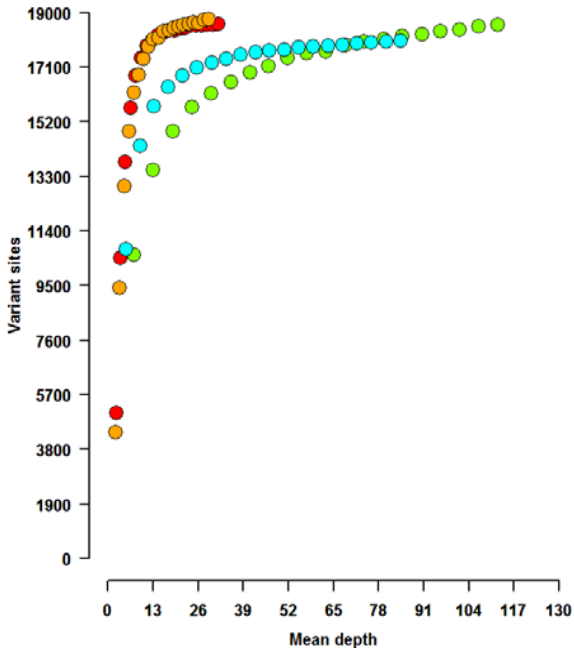

**B. Variants (platform-specific filters applied)**

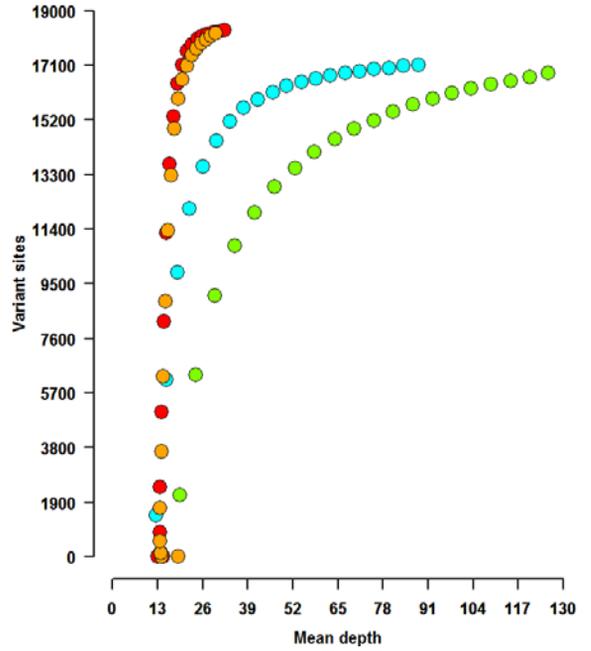

**C. Novel variant rate (unfiltered)**

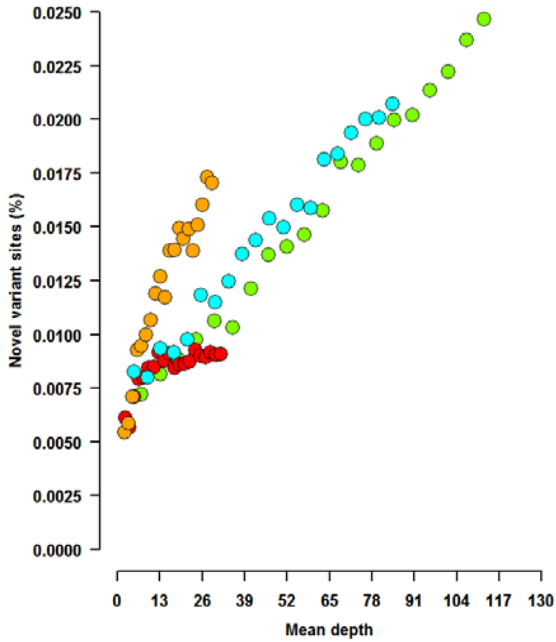

**D. Novel variant rate (platform-specific filters applied)**

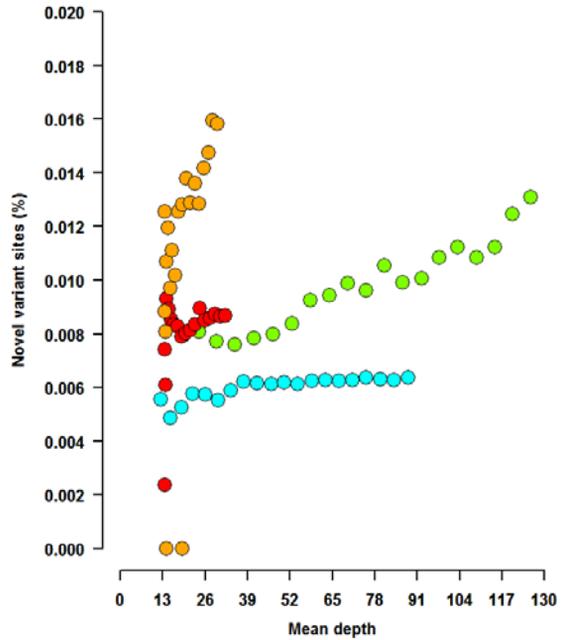

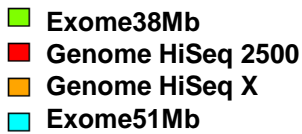

- Exome38Mb
- Genome HiSeq 2500
- Genome HiSeq X
- Exome51Mb

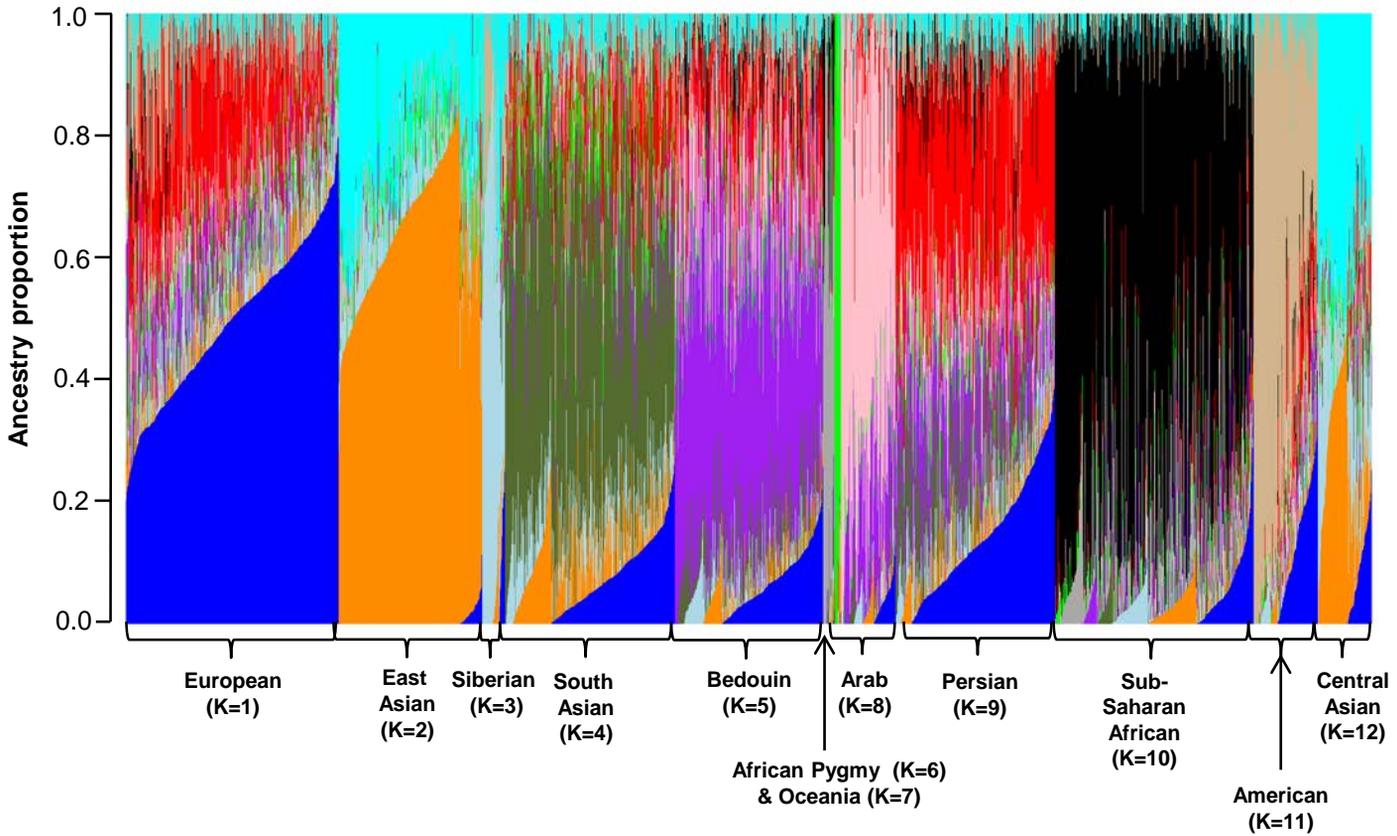

**Qatari, 1000 Genomes, and Human Origins individuals (n=5611), sorted by dominant ancestry cluster (K)**

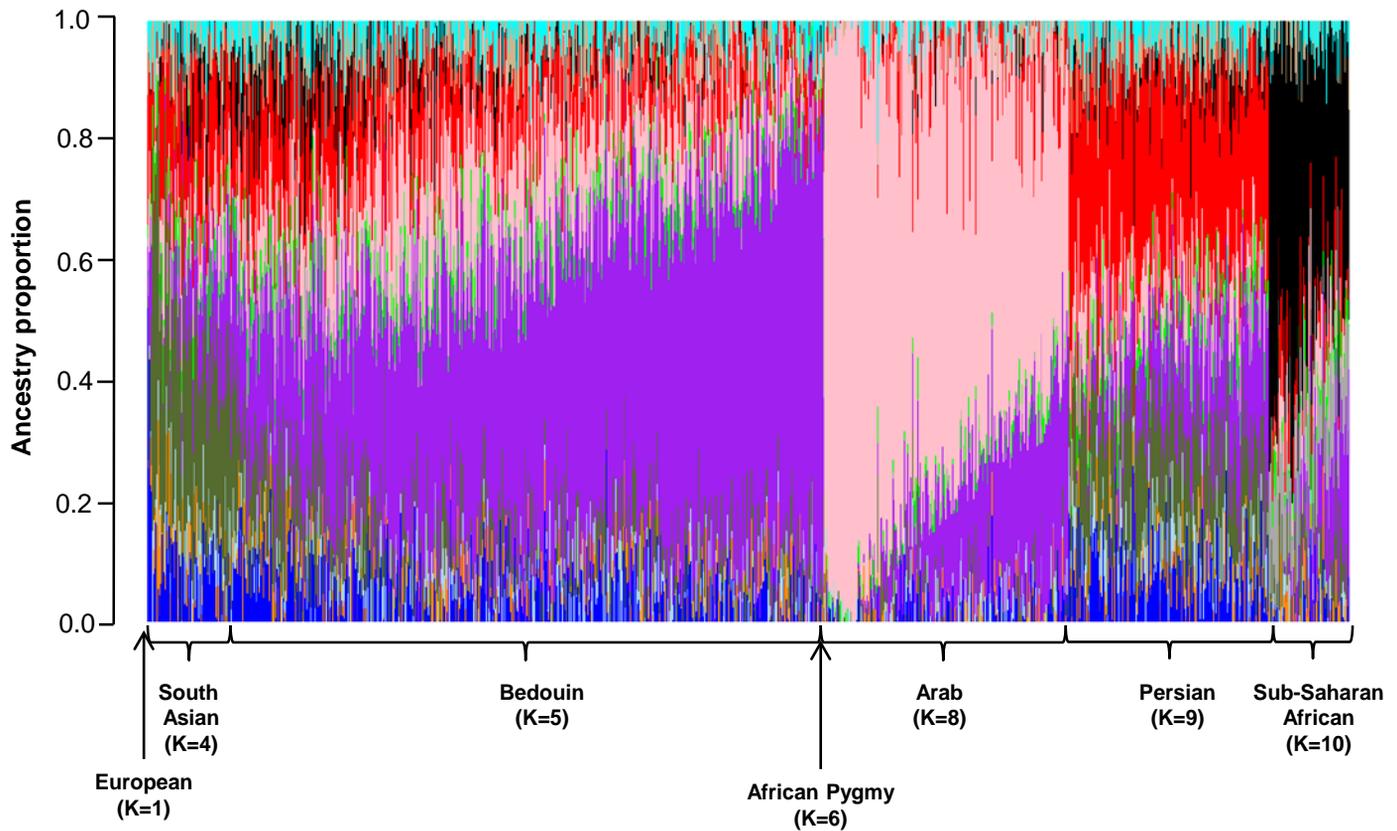

**Qatari individuals (n=1161), sorted by dominant ancestry cluster (K)**

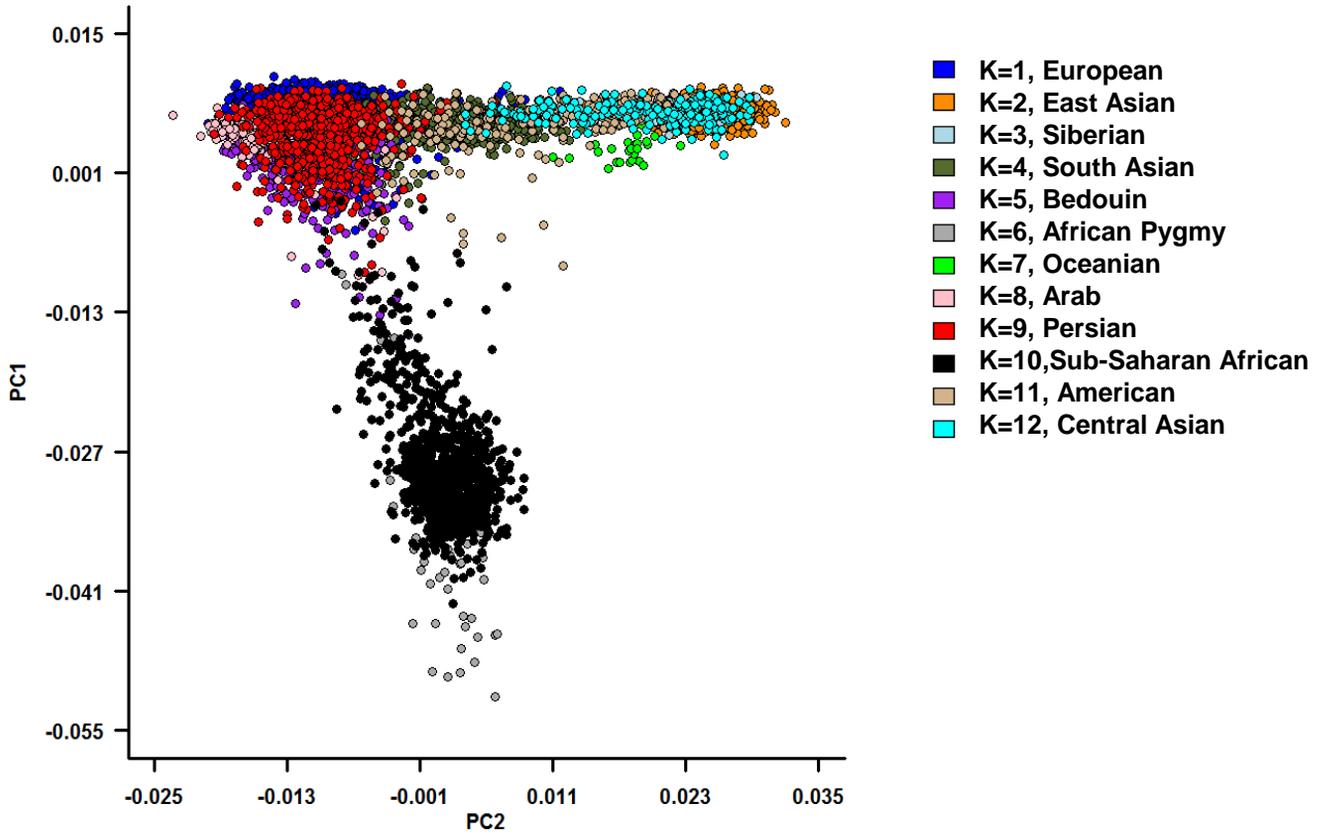

**A. PC1 and PC2 of genetic variation in Qatari, non-Qatari living in Qatar, 1000 Genomes Phase 3, and Human Origins (n=5611), color-coded by dominant ancestry cluster (K)**

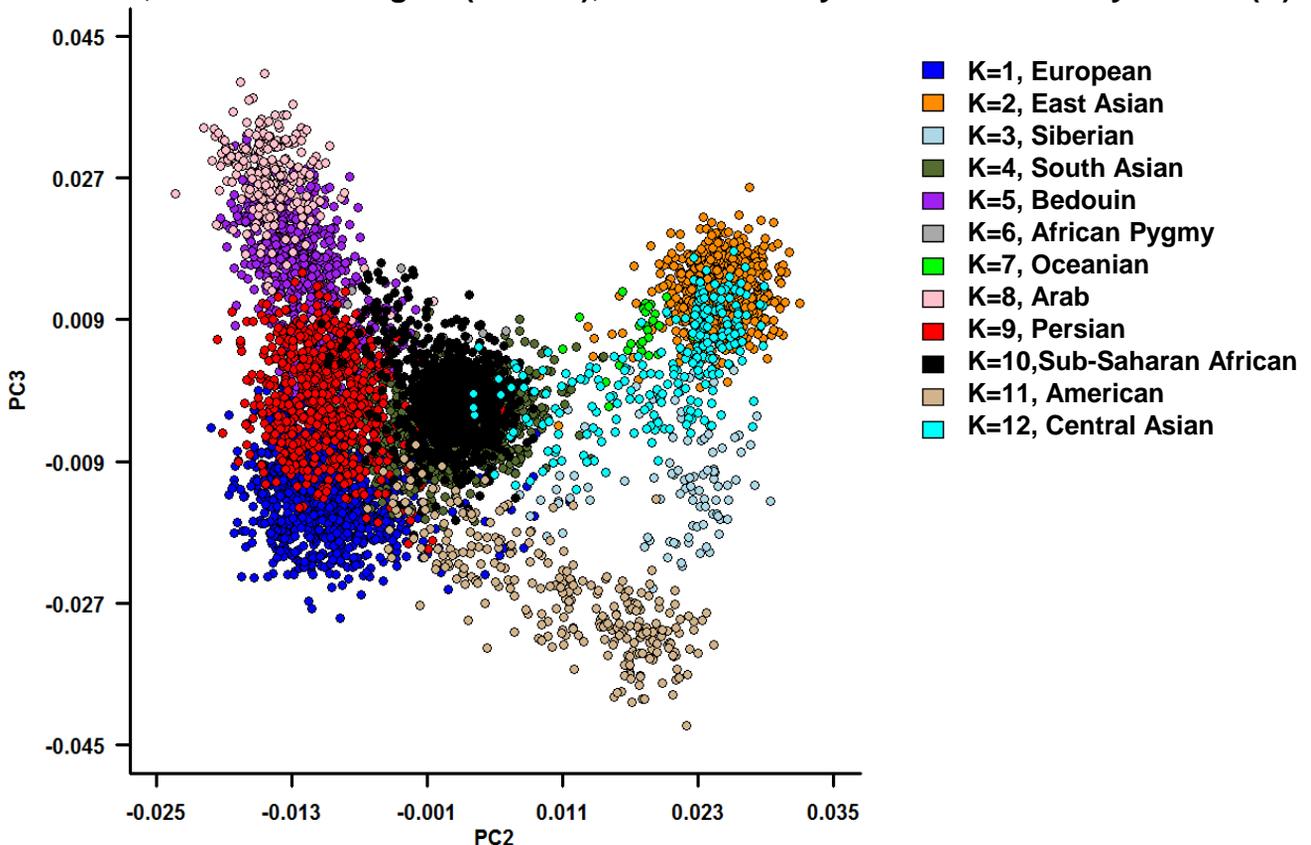

**B. PC2 and PC3 of genetic variation in Qatari, non-Qatari living in Qatar, 1000 Genomes Phase 3, and Human Origins (n=5611), color-coded by dominant ancestry cluster (K)**

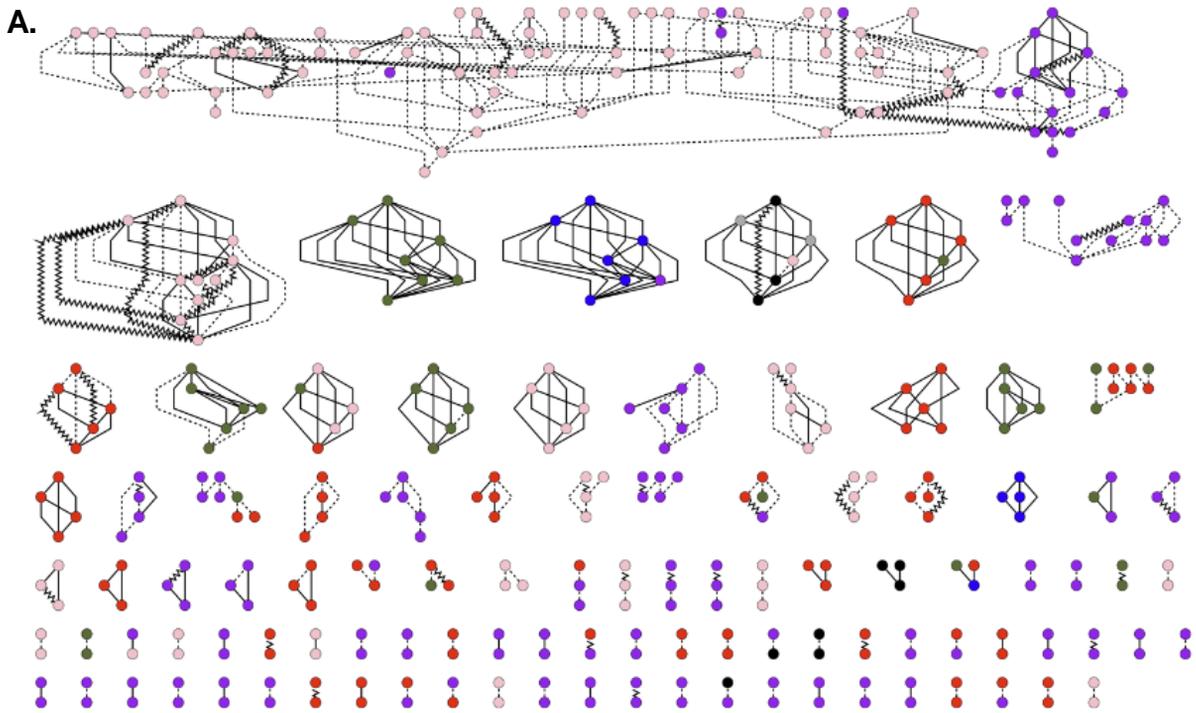
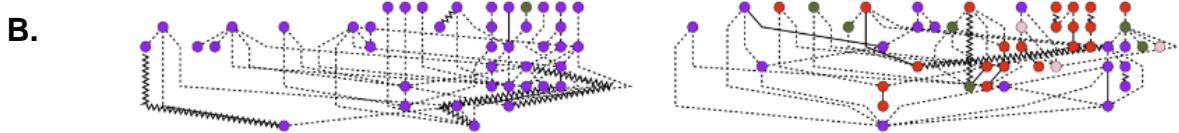

**Node color, based on ancestral cluster**

- K=1, European
- K=2, East Asian
- K=3, Siberian
- K=4, South Asian
- K=5, Bedouin
- K=6, African Pygmy
- K=7, Oceanian
- K=8, Arab
- K=9, Persian
- K=10, Sub-Saharan African
- K=11, American
- K=12, Central Asian

**Relationship degree**

- 1st degree
- 2nd degree
- 3rd degree